\documentclass[referee,useAMS,usenatbib,usegraphicx]{mn2e}
\usepackage{color}

\newcommand{\hii}{\mbox{H~{\sc ii}~}}
\newcommand{\oiii}{\mbox{[O~{\sc iii}]}~}
\newcommand{\sii}{\mbox{[S~{\sc ii}]}~}
\newcommand{\hei}{\mbox{He~{\sc i}~}}
\newcommand{\heii}{\mbox{He~{\sc ii}~}}
\newcommand{\niii}{\mbox{N~{\sc iii}}~}
\newcommand{\ciii}{\mbox{C~{\sc iii}}~}
\newcommand{\fiii}{\mbox{F~{\sc iii}}~}
\newcommand{\mgii}{\mbox{Mg~{\sc ii}~}}
\newcommand{\siiii}{\mbox{Si~{\sc iii}~}}
\newcommand{\nai}{\mbox{Na~{\sc i}~}}
\newcommand{\cai}{\mbox{Ca~{\sc i}~}}
\newcommand{\feii}{\mbox{Fe~{\sc ii}~}}
\newcommand{\fei}{\mbox{Fe~{\sc i}~}}

\title[Young  cluster NGC 1624]{A multi-wavelength census of stellar contents in  the young cluster NGC 1624}
\author[Jose et al.]{Jessy Jose $^{1}$\thanks{E-mail: jessy@aries.res.in}, 
A.K. Pandey$^{1}$, K. Ogura$^2$, D.K. Ojha$^3$,  B.C. Bhatt$^4$, 
\newauthor
M.R. Samal$^{1}$, N. Chauhan$^{1}$, D.K. Sahu$^4$ and  P.S. Rawat$^5$\\\\
$^1$ Aryabhatta Research Institute of observational sciencES (ARIES), Manora Peak,
Naini Tal, 263129, India\\
$^2$ Kokugakuin  University, Higashi, Shibuya-ku, Tokyo, 150-8440, Japan\\
$^3$ Tata Institute of Fundamental Research, Mumbai (Bombay), 400 005, India\\
$^4$ CREST, Indian Institute of Astrophysics, Koramangala, Bangalore, 560 034, India\\
$^5$ Department of Physics, D.S.B. Campus, Kumaun University, Naini Tal, India\\ }

\begin{document}

\date{}

\pubyear{2010}

\maketitle

\label{firstpage}

\begin{abstract}

We present a  comprehensive multi-wavelength analysis of the  young cluster NGC 1624 
associated with the \hii region Sh2-212 using optical $UBVRI$ photometry, optical 
spectroscopy and GMRT radio continuum mapping along with the  near-infrared (NIR)
$JHK$ archival data.  From optical observations of the massive stars, reddening $E(B-V)$ and  
distance to the cluster are estimated to  be 0.76 - 1.00  mag and $6.0 \pm 0.8$ kpc, 
respectively. Present analysis yields a spectral class of  O6.5V for the  main  ionizing
source  of the region  and the maximum post-main-sequence age of the cluster is estimated as $\sim$ 4 Myr.  
Detailed physical properties of the young stellar objects (YSOs)  
in the region are analyzed using a combination of optical/NIR colour-colour and colour-magnitude diagrams.
The distribution of YSOs in $(J-H)/ (H-K)$ NIR colour-colour diagram shows that a majority of them  
have  $A_V$ $\le$ 4 mag. However, a  few YSOs show $A_V$ values higher than 4 mag.  Based  on the
NIR excess characteristics, we identified 120  probable  candidate YSOs in  this region which yield a disk 
frequency of $\sim$ 20\%. However, this should be considered as a lower limit. These YSOs are found to 
have an age spread of $\sim$ 5 Myr with a median age of $\sim$ 2-3 Myr and a  mass range of $\sim$ 0.1 - 
3.0  $M_\odot$. A significant number of YSOs are located close to the cluster centre  and we detect an 
enhanced density of reddened YSOs located/projected close to  the molecular clumps  detected by 
Deharveng et al. (2008) at the  periphery of NGC 1624. This indicates that the YSOs located within 
the cluster core are relatively older in comparison to those located/projected near the clumps.
From the radio continuum flux, spectral class of  the ionizing source of the ultra-compact \hii (UC \hii) region 
at the periphery of Sh2-212 is estimated to be $\sim$ B0.5V. From optical data, slope of the mass function 
(MF) $\Gamma$,  in the mass range $1.2 \le M/M_{\odot}<27$  can be represented by a  single power 
law with a slope -1.18 $\pm$ 0.10, whereas the NIR data in the mass range $0.65 \le M/M_{\odot}<27$ 
yields $\Gamma$ = -1.31 $\pm$ 0.15.  Thus the MF  agrees fairly with the Salpeter value. The slope 
of the $K$-band luminosity function (KLF) for the cluster is found to be 0.30 $\pm$ 0.06 which is 
in agreement with the values obtained for other  young clusters. 

\end{abstract}

\begin{keywords}

stars: formation $-$ stars: luminosity function, mass function $­-$ stars:
pre$-$main$-$sequence $-­$ open clusters and associations: individual: NGC 1624.

\end{keywords}

\section{Introduction}
\label{intro}

The study of the  star formation process and the origin of stellar initial mass function (IMF),  
defined as the  distribution of stellar masses at the time of birth, are key issues in 
astrophysics.  Since majority of  stars tend to form in clusters or groups, young star 
clusters are considered to be the fundamental units of star formation (Lada \& Lada 2003).  
Young star clusters are useful tool to study the IMF as they  contain statistically 
significant number of young stars of rather similar age spanning a wide range of masses. 
Since  these objects are not affected by the dynamical evolution as the ages of these objects 
are significantly less in comparison to their dynamical evolution time, the present day mass 
function (MF)  of these objects can be considered as the IMF.  However, a recent study by Kroupa 
(2008) argues that even in the youngest clusters, it is difficult to trace the IMF, as clusters 
evolve rapidly and therefore eject a fraction of their members even at a very young age.

In the last decade, there have been a large number of studies in great detail in several
young clusters within 2 kpc of the Sun investigating these issues (e.g., Lada \& Lada 2003, 
Pandey et al. 2008, Jose et al.  2008). Although the theoretical 
expectation is that the IMF of a cluster should depend on the location, size, metallicity, 
density of the star forming environment and other conditions such as temperature or pressure 
(Zinnecker 1986; Larson 1992; Price \& Podsiadlowski 1995),  for clusters  located 
within 2 kpc, there is no compelling evidence for variation in the stellar IMF above the solar 
mass (e.g. Meyer et al. 2000; Kroupa 2002; Chabrier 2005).

With the aim of understanding the star formation process and IMF in/around young star clusters, 
we selected an young cluster NGC 1624  ($\alpha_{2000}$ = $04^{h}40^{m}38^{s}.2$; $\delta_{2000}$ 
= $+50^{\circ}27^{\prime}36^{\prime\prime}$;  l=155.36;  b=+2.62)  associated with the bright 
optical \hii region Sh2-212 (Sharpless 1959). A  colour composite image  using the bands $B$, 
blue; \oiii, green; and \sii, red  for an area $ \sim 10\times10$ arcmin$^2$   centered at   NGC 1624  
is shown in   Fig. \ref{cfht} (left panel), where the cluster seems to be embedded in the \hii region.
The cluster is located significantly above 
the  formal galactic plane ({\it Z} $\sim$  250 pc) for an estimated distance  of 6.0 kpc 
 (cf. Sect. \ref{distance}).  The kinematic and spectrophotometric distances to NGC 1624 vary 
from 4.4 kpc (Georgelin \& Georgelin 1970) to 10.3 kpc (Chini \& Wink 1984).  An IRAS point 
source (IRAS 04366+5022) with colours similar to that of the  ultra-compact \hii (UC\hii) region 
(Wood \& Churchwell 1989) is located at the periphery of Sh2-212. The molecular gas distribution 
of this region was mapped by CO observations (Blitz et al. 1982; Leisawitz et al. 1989; 
Deharveng et al. 2008).

Particularly,  Deharveng et al. (2008) studied  the region using $J = 2-1$ lines of $^{12}$CO and $^{13}$CO 
and reported a bright and thin semi-circular structure of  molecular gas (in the velocity 
range -34.0 kms$^{-1}$ to -32.7 kms$^{-1}$) in $^{13}$CO at the rear side of Sh2-212  along 
with  a filamentary structure (-36.8 kms$^{-1}$ to -35.9 kms$^{-1}$)  extending from southeast 
to northwest. The semi-circular ring  itself contains several molecular clumps, the most 
massive of which (-36.1 kms$^{-1}$ to -35.1 kms$^{-1}$) contains a massive young stellar 
object (YSO) which is the exciting source of the  associated UC\hii region (see Fig. 1). 
They concluded that Sh2-212 is a good example of massive-star formation triggered via the 
collect and collapse process. They also reported the flow of ionized gas and suggested that 
this may be the indication of `Champagne flow' towards the north of Sh2-212. A careful view
of Fig. 1 (right panel) reveals that the central region of  NGC 1624 is relatively devoid
of gas and  dust, whereas the outer regions, particularly east, south-east and west seem to be
obscured by molecular gas. However, it is to be noted that the semi-circular structure  
containing clumps is located at the rear side of the cluster. 

The present study is an attempt to understand the  stellar content, young stellar  population 
and the form of IMF/ $K$-band luminosity function (KLF) of the cluster NGC 1624 associated 
with Sh2-212  using our  optical and radio continuum observations along with the 
near-infrared (NIR) archival  data. In Sections 2 and 3, we describe the observations, data reductions
and archival data used in the present work. Sections 4 to 8 describe various cluster parameters
and young stellar properties derived  using optical, NIR and radio continuum data. 
Sections 9 and 10  describe the IMF and KLF of the region and in section 11 we have summarized 
the results.

\section{OBSERVATIONS AND DATA REDUCTIONS}

In the following sections  we describe 
the observations and data reductions carried out in order  to have a detailed study of NGC 1624.

\subsection{Optical CCD Photometry}
\label{obs}

The CCD $UBVRI$ observations of NGC 1624 were carried out using   Hanle Faint Object Spectrograph 
and Camera (HFOSC) of the 2-m Himalayan  Chandra Telescope (HCT) of Indian Astronomical Observatory 
(IAO), Hanle, India on 2004 November 3. The 2048 $\times$ 2048 CCD with a plate scale of 0.296 arcsec 
pixel$^{-1}$ covers an area of $\sim$ 10$\times$10 arcmin$^2$ on the sky.  We  took short and long
exposures  in all filters to avoid saturation of bright  stars. PG 0231 field from  Landolt (1992) 
was observed to determine atmospheric extinction as well  as to photometrically calibrate the  CCD 
frames on  the same night. The log of observations is tabulated in Table \ref{obslog}.

The CCD  frames were bias-subtracted  and flat-field  corrected in the standard manner using various 
tasks available under IRAF\footnote{IRAF is distributed by National Optical Astronomy Observatories,
USA}. Aperture photometry was done for the standard stars of PG 0231 field and the following 
calibration equations were derived using a least-squares linear regression:\\

\noindent
$(U-B) = (1.269\pm 0.020) (u-b) - (2.617\pm 0.026)$,\\
\noindent
$(B-V)=(0.915\pm 0.016) (b-v) - (0.284\pm0.012)$,\\
\noindent
$(V-R) = (1.056\pm 0.013) (v-r) - (0.011\pm0.010)$,\\
\noindent
$(V-I) = (1.022\pm 0.009) (v-i) + (0.188\pm0.008)$,\\
\noindent
$V = v+(0.024\pm 0.011) (V-I) - (0.495\pm0.013)$,\\

where,  $u,b,v,r,i$ are the instrumental magnitudes corrected for the atmospheric extinctions 
and $U,B,V,R,I$ are the standard magnitudes. The standard deviations of the  residuals, $\Delta$,
between standard and  transformed $V$ magnitudes, $(U-B)$, $(B-V)$, $(V-R)$ and $(V-I)$ colours of 
standard stars were 0.020, 0.045, 0.018, 0.014 and 0.021  mag, respectively. Different frames of 
the cluster region having same exposure  time and observed with the same  filters were averaged. 
Photometry of cleaned frames was carried out using the DAOPHOT-II (Stetson 1987) profile-fitting 
software.

We repeated the observations of NGC 1624  in $V$ and $I_c$ filters to get deeper photometry on  
2006 December 12 using   the 104-cm Sampurnanand Telescope  (ST)  of Aryabhatta Research Institute 
of observational sciencES (ARIES), Naini Tal, India. Log of the observations is given in Table 
\ref{obslog}.  The 2048 $\times$ 2048 CCD with a plate scale of 0.37 arcsec pixel$^{-1}$  covers a 
field of $\sim 13\times13$ arcmin$^2$ on the sky. To improve the  signal to noise  ratio (S/N), 
the observations were carried out in binning mode of $2\times2$ pixel. Secondary standards from  
the HCT observations were  used to  calibrate  the  data taken with ST. A combined photometry 
catalog is made using these two observations and this catalog has  typical  photometric errors
of the order of $\sim$  0.01 mag at brighter end ($V\sim$ 15), whereas the errors increase towards 
the fainter end ($\sim$ 0.04 at $V$ $\sim$ 21).  The catalog is available in electronic form and 
a sample table is  given in Table \ref{optdata}. 

In order to check the accuracy of the present photometry, we compared our photometry  with the 
$UBV$ photometry of 14 stars carried out by Moffat et al. (1979). The mean  and standard deviation 
of the difference between Moffat's and our photometry in $V$, $U-B$ and $B-V$ are $0.008 \pm
0.006$, $0.005 \pm 0.015$ and  $0.004 \pm 0.006$, respectively, suggesting that the  two photometries
are in good agreement. 
 
To study  the luminosity function (LF)/MF, it  is  necessary to  take into  account  the incompleteness
of  the present  data that could  occur due to various factors (e.g., crowding of the stars). We used  
ADDSTAR  routine of DAOPHOT-II to determine the completeness factor (CF). The procedure has been 
outlined in  detail in our earlier work (see e.g., Pandey et al. 2001). Briefly, we randomly added 
artificial stars to both $V$ and $I$ images taken with ST in such a way that they have similar 
geometrical locations but differ in $I$ brightness according to mean $(V-I)$ colour ($\sim 1.5$ mag) 
of the data sample.  Luminosity distribution of artificial stars was chosen in such a way that more 
number of stars were inserted towards the fainter magnitude bins. The frames were reduced using the 
same procedure used for the original frames. The ratio of the number of stars recovered to those added
in each magnitude interval gives the CF as a function of magnitude. Minimum value of the CF of the 
pair (i.e., $V$- and $I$-bands )  for the cluster region and field region (outside the cluster region), 
given in Table \ref{cf_opt}, is used to correct the data incompleteness. 

\subsection {Spectroscopic observations}

Low resolution optical spectroscopic observations of  4 optically bright sources  of NGC 1624 were made 
using HFOSC of HCT. The log of observations is given in Table \ref{obslog}. The  spectra in the 
wavelength range  3800-6840 $\AA$ with a dispersion of 1.45  $\AA$  pixel$^{-1}$ were obtained 
using low  resolution grism 7 with a slit having width  2$^{\prime\prime}$. One-dimensional spectra 
were  extracted from the  bias-subtracted and flat-field corrected images using the optimal 
extraction method in IRAF. Wavelength calibration of the spectra were done using FeAr and FeNe 
lamp sources. Spectrophotometric standard (Feige 110)  was observed on 2006 September 08 and  flux 
calibration  was applied to the star  observed on the same night. 

\subsection{Radio Continuum Observations}
Radio continuum observations at 1280 MHz were carried out on 2007 July 17 using the Giant Metrewave 
Radio Telescope (GMRT), India. GMRT has a `Y' shaped  hybrid configuration of 30 antennae, each of 
45 m diameter.  Details of the GMRT antennae and their configurations can be found in Swarup et al. (1991).
For the observations, the primary flux density calibrators used were 3C48 and 3C286. NRAO Astronomical 
Image Processing System (AIPS) was used for the data reduction. The data were carefully checked for
radio frequency  interference or other problems and suitably edited.  Self calibration was carried 
out to remove the residual effects of atmospheric and ionospheric phase corruptions and to obtain 
the improved maps.

\section {Archival data}

\subsection {Near-infrared data from 2MASS}

NIR $JHK_s$ data for  point sources within a radius of 10 arcmin around NGC 1624  have been obtained 
from Two Micron All Sky Survey (2MASS) Point  Source   Catalog (PSC) (Cutri et al. 2003).  To improve 
 photometric accuracy, we used photometric quality flag (ph$\_$qual = AAA) which gives a S/N $\ge$ 
10 and a  photometric uncertainty $ <$ 0.10 mag. This selection criterion ensures best quality detection 
in terms of photometry and astrometry as given on the  2MASS website\footnote 
{http://www.ipac.caltech.edu/2mass/releases/allsky/doc/}.  The $JHK_s$ data  were transformed  from 2MASS  
system  to the California Institute of Technology (CIT) system using   the relations given  by Carpenter 
(2001).  We used this data set to calibrate the NIR  archival data from Canada-France-Hawaii  Telescope 
(CFHT) (see Sect. \ref{cfhtdata}) and also to produce the radial density profile   of 
NGC 1624 (see Sect. \ref{rd}). 

\subsection {Near-infrared data from CFHT}
\label{cfhtdata}

NIR data for the region were obtained from  the Canadian  Astrophysical Data Centre's 
(CADC) archive program.  The NIR observations of the region were taken on 2002 October 20 
(PI: L. Deharveng) using the instrument CFHT-IR at the 3.56-m CFHT. The 1024 $\times$ 1024 
pixel HgCdTe detector with a plate scale of  0.211 arcsec/pixel was used for the observations.  
The  catalog by Deharveng et al. (2008) lists a total of 891 sources  in $JHK$ bands. Since 
our aim was to study the KLF of the region, where the estimation of the completeness of the 
photometry (ref. Sect.  \ref {obs}) was necessary,  we  re-reduced the CFHT  observations. We used 
dithered images at 9 different locations having 10 frames at each position around  the UC\hii 
region  of this field. Flat frames and sky frames were made from the median combined object frames. 
The sky subtracted and flat field corrected dithered images in each band were aligned and then combined 
to achieve a higher S/N.  The final  mosaic image covers an area of $5^{\prime}.2 \times 5^{\prime}.2$ 
with the UC\hii region at the centre  and is shown in Fig. \ref{cfht}. 

Photometry of the processed images were obtained  using the DAOPHOT-II package in IRAF.  Since the 
region was crowded, we performed PSF photometry on the images.  The 2MASS counterparts  of the 
CFHT sources were searched within a match radius of 1 arcsec. The CFHT instrumental magnitudes 
were compared  to the selected 2MASS magnitudes to 
define a slope and zero point for the photometric calibration. The rms    scatter between the 
calibrated CFHT  and 2MASS data (i.e.,  $2MASS - CFHT$ data)  for the $J, H$ and $K$-bands were
0.07, 0.08  and 0.06, respectively. In order to check the photometric accuracy,  we compared our 
photometry with the photometry reported by Deharveng et al. (2008).  The average dispersion between 
these two samples was $\sim$ 0.1 mag in $JHK$ bands  with absolutely no  shift, which shows that the 
present  photometry is in agreement with the previous study. To ensure good photometric accuracy, we 
limited our sample with those stars having error $<$ 0.15 mag in all three bands and thus we
obtained photometry for  951  sources in  $J, H$ and $K$-bands.  Additional 31 sources  detected only
in the $H$ and $K$ bands ($J$ drop out sources) having error $<$ 0.15 mag are also included in our 
analysis. Data of three saturated sources  have been  taken from the 2MASS catalog.  The detection limits were 
19.0, 18.4 and 18.0 mag for $J$, $H$ and $K$-bands, respectively.  We combined the optical  
and NIR catalog  within a match radius of 1 arcsec and the final catalog used in the present 
analysis is available  in electronic form  
and a sample table is shown in Table \ref{optdata}.  We estimated the completeness limit of the data
using the  ADDSTAR routine of DAOPHOT-II. The procedure was the same as mentioned for the optical 
images (see Sect. \ref{obs}). Completeness was greater than 90$\%$  for magnitudes brighter than 17.0 and 
reduced to 80 $\%$ for the magnitude range 17.0 - 17.5 in $K$-band. We did not find any significant 
spatial variation of the completeness factor within the entire area of $5^{\prime}.2 \times 5^{\prime}.2$ 
and hence we used an average completeness  factor of the region for our analysis.  

\section{Structure of the cluster}

\subsection{Two dimensional surface density distribution}

The initial stellar distribution in star clusters may be governed by the  structure of parental 
molecular cloud and also how star formation proceeds  in the cloud (Chen et al. 2004, Sharma et al. 2006). 
Later evolution of the  cluster may then be governed by internal gravitational interaction among 
member stars and external tidal forces due to the Galactic disk or giant molecular clouds. 

To study the morphology of the cluster, we  generated isodensity contours for stars in $K$-band  from CFHT 
data and is shown in Fig. \ref{ssnd}. The contours are plotted above 3-sigma value of the 
background level as estimated from the control field. The star mark in Fig. \ref{ssnd} represents the 
location of the cluster centre (Sect. \ref{rd}). The surface density distribution of the CFHT data
reveals prominent sub-structures which seem to be distributed symmetrically around the cluster 
centre at a radial distance of $\sim$ 35 arcsec. Interestingly, these sub-structures are lying just 
inside the thin molecular layer shown in Fig.  \ref{cfht}. 

\subsection{Radial stellar surface density and cluster size}\label{rd}

The radial extent  of a cluster is one of the important parameters used to study the dynamical  
state of  the cluster. We used the star count technique  to  study the surface density distribution 
of stars in the  cluster region and to derive the radius of the cluster. To  determine   the  
cluster  centre,  we  used   the  stellar density distribution of  stars   in a $\pm$  30 pixel 
wide  strip along  both X  and Y  directions around  an   eye estimated centre. The  point of
maximum density obtained  by fitting a Gaussian  curve was   considered as  the  centre of  the 
cluster.   The coordinates of  the cluster centre were found to be  $\alpha_{2000}$ = 
$04^{h}40^{m}38^{s}.2 \pm 1^{s}.0$; $\delta_{2000}$ = $+50^{\circ}27^{\prime}36^{\prime\prime} \pm 15^{\prime\prime}$.

To investigate the radial structure of the cluster, we derived the radial density profile (RDP)  
using the ST observations for  $V \le $  20  mag  and 2MASS $K_s$-band data ($K_s \le$ 14.3 mag).  
Sources were counted in concentric annular rings of 30 arcsec width around the cluster centre and 
the counts were normalized  by the area of each annulus.  The  densities thus obtained are plotted as 
a function of radius in Fig. \ref{rad}, where, one arcmin at the distance of the cluster (6.0 kpc, 
cf. Sect. \ref{distance}) corresponds to $\sim$ 1.8 pc. The upper and lower panels  show the RDPs 
obtained from optical  and 2MASS $K_s$-band data, respectively. The error bars are derived assuming 
that the number of stars in each annulus follows Poisson statistics.

Radius of the cluster $(r_{cl})$ is defined as the point where the cluster stellar density merges  
with the  field stellar  density.  The horizontal  dashed line  in Fig. \ref{rad} shows  the field 
star density. For the optical RDP, the field star density is determined from the  corner of our 
optical CCD image, whereas for the NIR RDP, the field star density is determined from an area which 
is 10 arcmin  away from the cluster centre. The error limits in the field density distribution 
are shown using dotted lines.  

To parametrize the RDP, we fitted the observed RDP with the empirical model of King (1962) which is 
given by
\begin{equation} \hspace{20mm}{\rho (r) = {{\rho_0} \over
\displaystyle {1+\left({r\over r_c}\right)^2}}}
\end{equation} where  $r_c$ is  the core  radius at  which the surface density $\rho(r)$ becomes  
half of  the central  density, $\rho_0$. The  best fit  to the observed RDPs  obtained by a $\chi^2$
minimization technique  is shown in Fig. \ref{rad}. The core  radii thus estimated  from optical 
and NIR RDPs are 0.50 $\pm$ 0.06  and 0.48 $\pm$ 0.05 arcmin, respectively.  Within errors, the King's
profile (Fig. \ref{rad}, solid curve) seems to be  merging with  the background field at $\sim$ 
2.0 arcmin both for the  optical and 2MASS data. Hence, we assign a radius of  2.0 arcmin  for NGC 1624. 
Here we would like to  point out that the core radius and boundary of  the cluster are estimated 
assuming a spherically symmetric distribution of  stars within the cluster. This approach is frequently
used to estimate the extent of a cluster. 

\section{Analysis of optical data} 

\subsection{Reddening in the cluster}
\label{reddening}

To study  the nature of the  extinction law  towards  NGC 1624, we used two-colour diagrams (TCDs) as  
described by Pandey et al. (2003). The  TCDs of the form of  ($V-\lambda$)  versus ($B-V$), where  
$\lambda$ is  one   of  the  broad-band filters ($R,I,J,H,K,L$),  provide  an effective method for 
separating the influence of normal extinction produced by the diffuse interstellar medium from that 
of the abnormal extinction arising within  regions having a peculiar  distribution of dust sizes 
(cf. Chini \& Wargau 1990; Pandey et al. 2000).  The ${E(V-\lambda)}\over  {E(B-V)}$ values in  
NGC 1624 are estimated  using the procedure as described in Pandey et al. (2003).  The slopes of 
the distributions $m_{cluster}$  are found to  be identical to the normal values as given in Pandey 
et al. (2003).  Thus we adopt a normal reddening law ($R_V=3.1$) for NGC 1624.

In the absence of spectroscopic observations, the interstellar extinction  $E(B -− V)$ towards the 
cluster region can be estimated using the $(U −- B )/(B -− V )$ colour-colour (CC) diagram. The CC
diagram of NGC 1624 ($r \le 2^\prime$) is presented in Fig. \ref{ubbv},  where, continuous curves 
represent the empirical  zero-age-main-sequence (ZAMS) locus by Girardi et al. (2002). The ZAMS 
locus is reddened by $E(B-V)$  = 0.76 and 1.00 mag  along the normal reddening vector 
(i.e., $E(U  - B) /E(B - V )$ = 0.72).  Fig. \ref{ubbv} indicates that majority of the $O-A$ type 
stars  have $E(B - V)$  in the range  of  0.76 -  1.00 mag.  The stars lying within the reddened 
ZAMS may be   probable   members of NGC 1624. Using $K/ (J-K)$ colour-magnitude  diagram (CMD), 
Deharveng et al. (2008) have also reported  $A_V \sim 3$ mag for the whole region. 
A careful inspection of the CC diagram indicates the presence of further reddened population which could be 
the probable background population of the region. The
theoretical ZAMS, shown by dashed line, is further shifted to match the reddened sequence. 
The $E(B - V)$ value for the background population comes out to be $\sim$ 1.15 mag.

Reddening of  individual stars having spectral types  earlier than A0 have  also been computed  by 
means  of  the reddening  free index $Q$ (Johnson $\&$ Morgan 1953).  Assuming a normal reddening law 
we can construct a reddening-free parameter index $Q = (U-B) - 0.72\times (B-V)$. For stars earlier 
than A0, value of $Q$ will be $<$ 0. For main-sequence (MS) stars, the intrinsic $(B-V)_0$ colour and
colour-excess can be obtained from the relation $(B-V)_0 = 0.332\times Q$ (Johnson 1966; Hillenbrand 
et al. 1993) and $E(B-V) = (B-V) - (B-V)_0$, respectively. The individual reddening of the massive 
stars down to  A0 spectral class within NGC 1624 ($r \le 2^\prime$) are found to vary in the range  
$E(B-V)$ $\simeq$  0.76 - 1.05 mag implying the presence of differential reddening within the cluster. 
The $A_V$ values thus calculated for stars up to A0 spectral class have been given in Table \ref{optdata}. 
Assuming the standard deviation of the residuals (cf. Sect. \ref{obs}) as typical errors 
in  photometry, we estimate a typical error in estimation of $E(B-V)$  as $\sim$ 0.05 mag.

\subsection {Spectral classification  of the bright sources in NGC 1624}
\label{slitspec}

We carried out low  resolution spectroscopy of  four optically bright sources within   2 arcmin radius 
of NGC 1624.  These sources are referred  as M2,  M4, M9 and M8 (see Fig. 6 of Deharveng et al. 2008). 
The brightest source M2 is  the probable  ionizing source of Sh2-212 (Moffat et al. 1979).  This star 
was  identified as an emission line star of class O5e by Hubble (1922). Moffat et al.  (1979) classified
this object as O5.5V star, whereas Chini \& Wink (1984) classified it as O6I type star.  To determine 
the  spectral type of this star, we extracted   low-resolution,  one dimensional spectrum.  In the top
panel of Fig. \ref{spec}, we show the flux calibrated, normalized spectrum of the ionizing source M2  
with important lines identified and labeled.  Among the Balmer lines, $H{\alpha}$ and $H_{\beta}$ are 
relatively strong in emission compared to  $H{\gamma}$, which is weak in emission. The $H{\delta}$ and 
$H{\epsilon}$ are in absorption. The other lines found in emission are   \heii $\lambda$  4686  and
\ciii $\lambda\lambda$   4647-50. 

In the case of early type stars, the ratio of \hei $\lambda$ 4471/\heii  $\lambda$ 4542  is a primary 
indicator of the spectral type.  This ratio is found to vary from   less than 1 to 1 and greater than 1 
as we move from O5 to O7 and later types.  The presence of strong \heii  $\lambda$ 4542 in absorption  
which is often accompanied by weak  \niii $\lambda\lambda$ 4634-42 emission indicate a MS  luminosity 
class denoted by ((f)). The absorption strength of \heii $\lambda$  4686 weakens while \niii emission 
strength increases in intermediate luminosity classes, denoted by (f) category. Finally, the Of super 
giants show both \heii and \niii in  strong emission (Walborn \& Fitzpatrick 1990). 

The ratio of \hei $\lambda$ 4471/\heii  $\lambda$ 4542 for M2 is found to be   (i.e., Log EW = Log 
(EW(\hei $\lambda$ 4471)/EW(\heii $\lambda$ 4542)) -0.15,  implying that this star is likely to be of
spectral type earlier to O7.  Following Conti \& Alschuler (1971) we assign  O6.5 $\pm$ 0.5 spectral 
type to this star. The weak nature  of \niii $\lambda\lambda$ 4634-42 indicates that this star is likely 
to be in MS. Thus we assign a spectral class of  O6.5 $\pm$ 0.5 V for the ionizing source of Sh2-212. 

The bottom panel of Fig. \ref{spec} shows the low resolution spectrum for the star  M4. The absence of 
\heii $\lambda$ 4200,  \heii $\lambda$ 4686   and  \mgii $\lambda$ 4481  indicates that the spectral 
class of M4 is between B1-B2 (Walborn \& Fitzpatrick 1990). The lack of spectral lines  \mgii  $\lambda$  
4481  and \siiii $\lambda$ 4552 rules out the possibility of it being an evolved star.   A comparison with 
the low resolution stellar spectra of Jacoby et al. (1984) and Walborn \& Fitzpatrick (1990) suggests 
this star as a spectral class  of B1.5 $\pm$ 0.5 V.

 The  reddening slope E(B-V)/E(U-B)  has also been obtained using the spectral types of the M2 (06.5V ) 
and M4 (B1.5V) stars. The value  of the slope using the intrinsic values from  Koorneef (1984) / Johnson (1966)
comes out to be  0. 86 / 0.83 and  0.75 / 0.73 for M2 and M4, respectively. The reddening slope for the B type 
star agrees well the value obtained in Sect. \S\ref{reddening}. We adopt a normal reddening law in the region 
as mentioned in  Sect. \S\ref{reddening} for further analysis of the data. 

We also extracted the low resolution  spectra (not shown here) for the stars M8 and M9. Presence of the 
spectral lines \nai  $\lambda$ 5893, \cai $\lambda$$\lambda$ 6122, 6162,  \feii $\lambda$ 6456  and the 
line strength of \fei, \cai $\lambda$ 6497 put these two stars in the mid F giant category  based on the 
spectral atlas given by Torres-Dodgen \& Weaver (1993) and Jacoby et al. (1984). 

\subsection{Optical colour-magnitude diagrams : Distance and age}
\label{distance}

The optical colour-magnitude diagrams (CMDs) are useful  to derive the cluster 
fundamental parameters such as age, distance etc. Fig. \ref{q} shows dereddened 
$V_0/(B-V)_0$ CMD for  probable cluster members (Sect. \ref{reddening}) lying within 
$r \le 2^{\prime}$  of NGC 1624. The stars having spectral  type earlier than A0 were
dereddened individually using  $Q$  method as discussed in Sect. \ref{reddening}. 
The stars labeled  as M2, M4, M8 and M9 (following the nomenclature by Deharveng 
et al. 2008) have spectroscopic observations as discussed  in Sect. \ref{slitspec}. 
The spectral class of the ionizing source (M2; see Sect. \ref{slitspec}) yields 
intrinsic distance  modulus of 14.05  which  corresponds to  a distance of 6.5 kpc, 
whereas the spectral class of M4 yields intrinsic  distance modulus of 13.8 which 
corresponds to  a distance of 5.8 kpc. The average distance from these two 
spectroscopically identified cluster members comes out to be 6.15 kpc.   We also 
calculated the individual distance modulus  of the remaining 12 probable MS stars (shown as
filled circles in Fig. \ref{q}). The intrinsic colours for each star were estimated using  the $Q$ method
as discussed in  Sect. \ref{reddening}. Corresponding $M_V$ values have been estimated 
using the ZAMS by Girardi et al. (2002). The average value of the intrinsic distance 
modulus obtained from the 14 stars (2 from spectroscopy and 12 from photometry) 
comes out to be 13.9 $\pm$ 0.3 which corresponds to 
a  distance of $6.0 \pm 0.8$ kpc. In  Fig. \ref{q} we have also plotted the theoretical 
isochrone of 2 Myr ($Z=0.02$; log age = 6.3) by Girardi et al. (2002), shifted for the 
distance modulus of  $(m-M_V)_0$ = 13.90  $\pm$ 0.3, which seems to be matching  well 
with the distribution of the probable MS members of the cluster.  Present distance 
estimate is in agreement with that obtained by Moffat et al. (1979; 6.0 $\pm$ 0.5 kpc), 
whereas Chini \& Wink (1984) have reported a distance of 10.4 kpc to NGC 1624. The  
distance estimates by Moffat et al. (1979) and Chini \& Wink (1984) were based  on the 
assumed spectral class of the ionizing source M2 (i.e., O5.5V and O6I, respectively). 
Here, it is worthwhile to mention  that the $M_V$ value for an O6V star in the literature 
varies significantly; e.g.,  $M_V$ = -5.5 (Schmidt-Kaler 1982) to -4.9 (Martins  et al. 2005).  
Hence, the distance estimation based on the O-type star  alone may not be reliable. However, 
the present distance estimation is carried out  using the O-type star as well as all  the 
probable members  earlier to A0  spectral type.   The kinematic distance (6.07 kpc) to the 
region derived by  Caplan et al. (2000) is in agreement with the present distance estimation. 
Since this cluster is located in the outer galactic disk, the possibility of a low metallicity 
for the region cannot be ruled out, which would imply bluer intrinsic colour for the members 
and hence a closer distance of NGC 1624. However, in the absence of any metallicity measurements 
towards this region, we have considered solar metallicity for the region and the distance of 
NGC 1624 is taken as 6.0 kpc for the present study.

The ages of young clusters are typically derived from the post-main-sequence evolutionary
tracks for the earliest members if significant evolution has occurred  and/or by fitting the 
low-mass contracting population with theoretical PMS isochrones. Since the most massive member of 
NGC 1624  seems to be a O6.5  MS star, the maximum age of the cluster  should be of the order of 
the MS life time of the massive star i.e., $\sim$ 4.4 Myr (Meynet et al. 1994).  In Fig. \ref{q} 
we have also shown the isochrone of 4 Myr age by Girardi et al. (2002), which suggests that 
the maximum  post-main-sequence age of the cluster could  be  $\sim$ 4 Myr. Stars which deviate  
significantly from the isochrone are likely 
field stars and are shown by open circles  in Fig. \ref{q}, which include stars M8 and M9.  
Spectroscopic observations of these two stars indicate that  they are of mid F giant spectral 
category (see Sect. \ref{slitspec}) and hence cannot be the cluster members at this assumed distance 
and age.  

$V/(V - I)$ CMD for the stars lying within the core of the  cluster ($r \le 0^\prime$.5)
is shown in Fig. \ref{cmd}a and CMD for the stars outside the core ($0^\prime.5 \le r \le 2^\prime$) 
is shown in Fig. \ref{cmd}b.  In order to  find out  the field star contamination 
in the cluster region, we selected a control field   having same area as that of the 
cluster from the corner of our CCD image.  $V/(V - I)$  CMD for the control  field is  
shown in  Fig. \ref{cmd}c. Assuming  $E(B-V)_{min} =0.76$ mag, $E(B-V)_{max}$ =1.0 mag 
and using the  relations $A_{V}=3.1\times  E(B-V)$; $E(V-I)=1.25\times E(B-V)$, we have  
plotted  theoretical isochrone of 2  Myr  by  Girardi et al. (2002) 
and   pre-main-sequence (PMS) isochrone of 0.5 and 5 Myr (Siess et al. 2000) in Fig. 
\ref{cmd}. It is evident from this figure that the  MS ($V \le $ 16.5) is rather free 
from field star contamination.  Although the CMDs  of the  cluster region  show a 
significant number of stars towards the right of the 2 Myr isochrone at $(V-I) >2.5 $  and 
$  V >18 $ mag, a comparison between the cluster and field regions clearly reveals the 
contamination due to field star population in  the CMD of the cluster region.  However, 
the $V/(V - I)$ CMD of the  core (Fig. \ref{cmd}a) reveals uncontaminated population 
of PMS stars having ages 0.5 - 5 Myr. 

As discussed in Sect. \ref{reddening}, there is indication for a population in the background of 
the cluster which is apparent in Figs. \ref{cmd}b and \ref{cmd}c. 
Assuming the average $E(B-V)$ = 1.15 mag, we estimate  that the distance of the  background population 
is $\sim$ 8 kpc. The study by Pandey et al. (2006) also indicates a background population at a distance 
of $\sim$ 8 kpc in the second galactic quadrant. 

\subsection {Emission from  ionized gas}
\label{ionized gas}

Fig. \ref{1280} shows GMRT radio continuum map of Sh2-212 at 1280 MHz made with a  resolution  of 
$\sim$ 4$^{\prime\prime}$.9 $\times$ 3$^{\prime\prime}$.2. In the  high resolution map, most of the
extended diffuse emission associated with the region appears quite faint. However,  a compact 
intense emission can be seen at the position of UC\hii region (04$^{\rm h}$40$^{\rm m}$27$^{\rm s}$.5, 
+50$^\circ$28$^\prime$28$^{\prime\prime}$)  located at the  periphery of Sh2-212 
and is marked using an arrow. The UC\hii region is  associated with the IRAS point source  
IRAS 04366+5022. The overall morphology of the map  agrees   well with that of our optical colour
composite image  shown in Fig. \ref{cfht}.  Fig. \ref{610} shows an  enlarged  version of the 
UC\hii region at 1280 MHz.  The integrated flux densities from the radio continuum contour maps 
for the evolved \hii region (i.e., Sh2-212) and UC\hii region are estimated to be 3.6 $\pm$
0.4 Jy and 16.5 $\pm$ 0.5 mJy, respectively.  Assuming the ionized regions to be spherically 
symmetric and neglecting absorption of ultraviolet radiation by dust inside the \hii region, 
the above flux densities  together with assumed distance, allow us to estimate the number of 
Lyman continuum photons (N$_{Lyc}$) emitted per second, and hence the spectral type of the 
exciting stars. Using the relation given by Mart\'{i}n-Hern\'{a}ndez et al. (2003)  for an 
electron temperature of 10000 K, we estimated  log N$_{Lyc}$ = 48.29  and log N$_{Lyc}$ = 45.96 
for the evolved \hii and UC\hii region, respectively,  which corresponds to MS spectral types of
$\sim$ O7 and $\sim$ B0.5, respectively (Vacca et al. 1996).  On the basis of optical spectroscopy, 
we estimated  spectral type of the ionizing source of   Sh2-212  as O6.5V (see Sect. \ref{slitspec})
which is in fair agreement with the above spectral type estimation from integrated radio continuum 
flux.  Using the spectral energy  distribution, Deharveng et al. (2008)  have found  that the source 
associated with the UC\hii region is a massive YSO of $\sim$ B0 type  ($\sim$ 14 $M_\odot$), which is 
in agreement with the spectral type $\sim$ B0 obtained in the present work.

\section{Analysis of  near-infrared  data}
\label{nir}

\begin{figure*} \centering 
\includegraphics[scale = .76, trim = 2 150 0 150, clip]{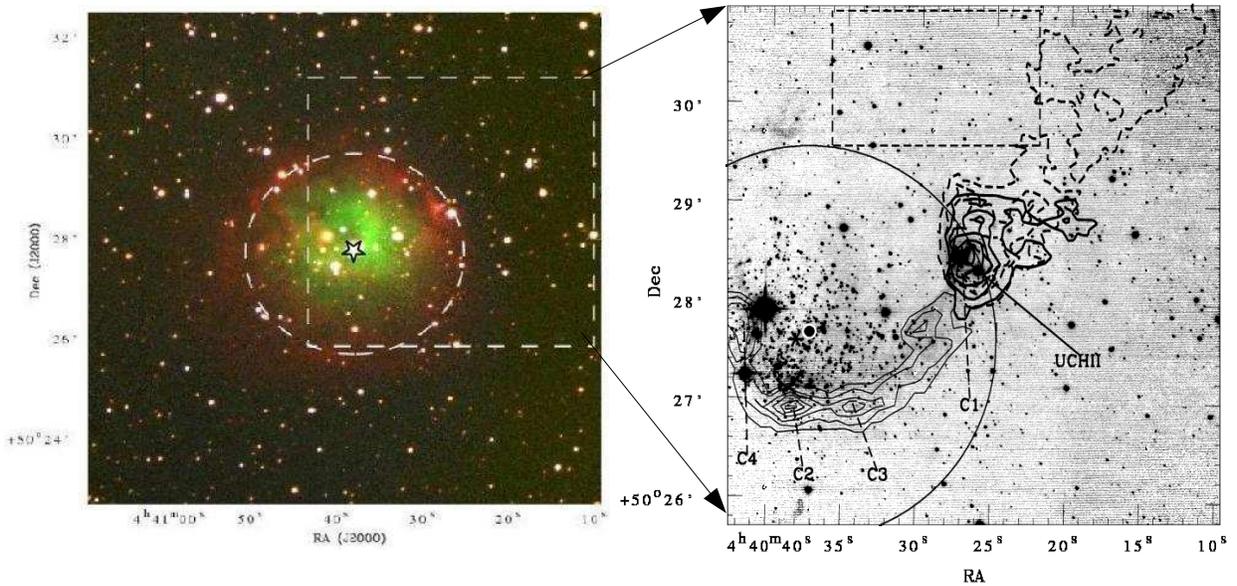}

\caption{$Left$: The colour composite image reproduced using the bands $B$, \oiii and 
\sii ($B$, blue; \oiii,  green; and \sii, red)  for an area  $\sim 10 \times 10$ arcmin$^2$  
around NGC 1624 (see the electronic version for the colour image). The  dashed line box  
represents the  $ 5^{\prime}.2 \times 5^{\prime}.2$ area of CFHT-$JHK$ observations 
(cf. Sect. \ref{cfhtdata}). 
The  star mark represents the cluster centre and the dashed circle  represents the  boundary of NGC 1624 
(cf. Sect. \ref{rd}).  $Right$: CFHT $K$-band mosaic image with a  field of view of 
$5^{\prime}.2 \times 5^{\prime}.2$  centered on the UC\hii region of Sh2-212. The white  circle  
represents the ionizing  source of Sh2-212  and the asterisk represents the  centre of NGC 1624. 
The contours represent the $^{13}$CO(2-1) emission map from Deharveng et al. (2008) in the 
velocity range between -34.0 kms$^{-1}$ to -32.7 kms$^{-1}$  (continuous thin contours),   
-36.1 $km s^{-1}$ to -35.1 $km s^{-1}$ (continuous thick contours) and -36.8 kms$^{-1}$ to 
-35.9 kms$^{-1}$ (dashed contours), respectively.  C1, C2, C3 and C4 are the 
molecular clumps  identified by Deharveng et al. (2008).  The partial circle shows 
$\sim$ 9.6 arcmin$^2$ section of the cluster (radius = 2$^\prime$; area = 12.6 arcmin$^2$). 
The control field region (cf. Sect. \ref{nir}) is represented by  the dashed line box. }

\label{cfht}
\end{figure*}

\begin{figure*}
\centering 

\includegraphics[scale =.5,trim=0 0 0 0, angle=-90, clip]{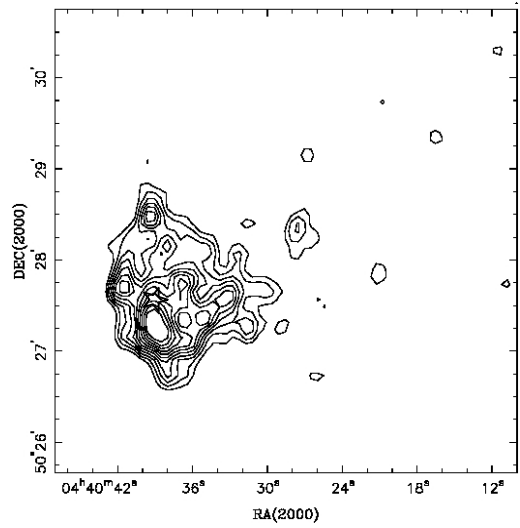}
\caption{The two dimensional stellar surface number density distribution obtained from the
CFHT $K$-band data using a grid size of $5^{\prime\prime} \times 5^{\prime\prime}$. The lowest contour  
is plotted at  3 times above the background level. The star mark represents the cluster centre.  }

\label{ssnd}
\end{figure*}

\begin{figure*}
\centering 
\includegraphics[scale =.5,trim=10 10 10 10, clip]{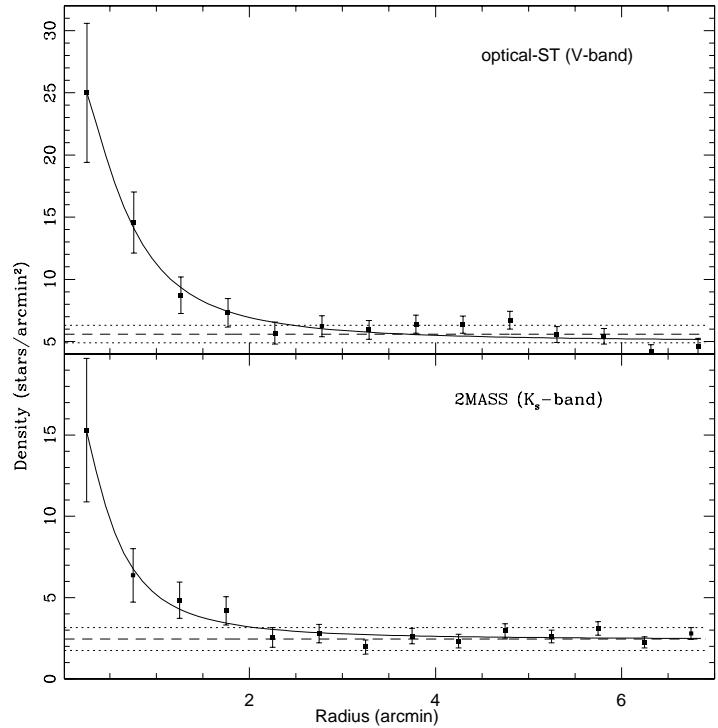}

\caption{Stellar  density as  a function  of radius  from  the adopted
cluster  centre for  the optical (upper panel)  and 2MASS (lower panel) data.  The solid  curve shows 
the least square  fit of the  King (1962) profile  to the observed
data points. The dashed  line represents the  mean density  level of the  
field stars and dotted lines are the error limits for the field star density.
The error bars  represent $\pm$  $\sqrt{N}$ errors.}
\label{rad}
\end{figure*}

\begin{figure*}
\centering 
\includegraphics[scale = .5, trim = 10 20 50 100, clip]{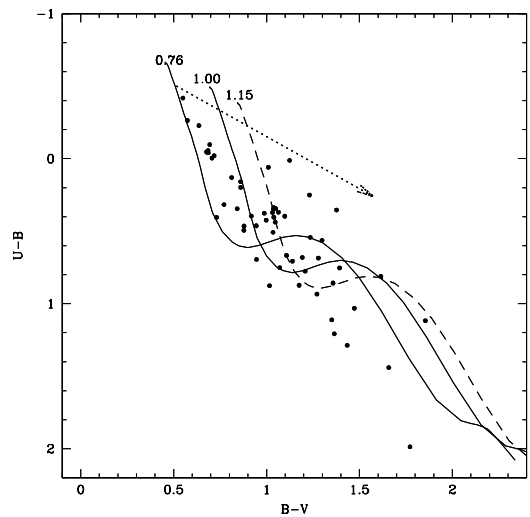}
\caption{$(U  - B)/(B  - V)$ colour-colour  diagram for the stars within  $r \le 2^\prime$ 
of NGC 1624. The continuous curves represent the ZAMS by Girardi et al. (2002) shifted 
along the reddening slope of 0.72 (shown as dotted line) for $E(B-V)$ = 0.76 and 1.00 mag,
respectively. The dashed curve represents the ZAMS reddened by E(B - V) = 1.15 mag to match  
the probable background population (see the text for details).
 }
\label{ubbv}
\end{figure*}

\begin{figure*}
\centering 
\includegraphics[scale = .8, trim = 0 0 0 0, clip]{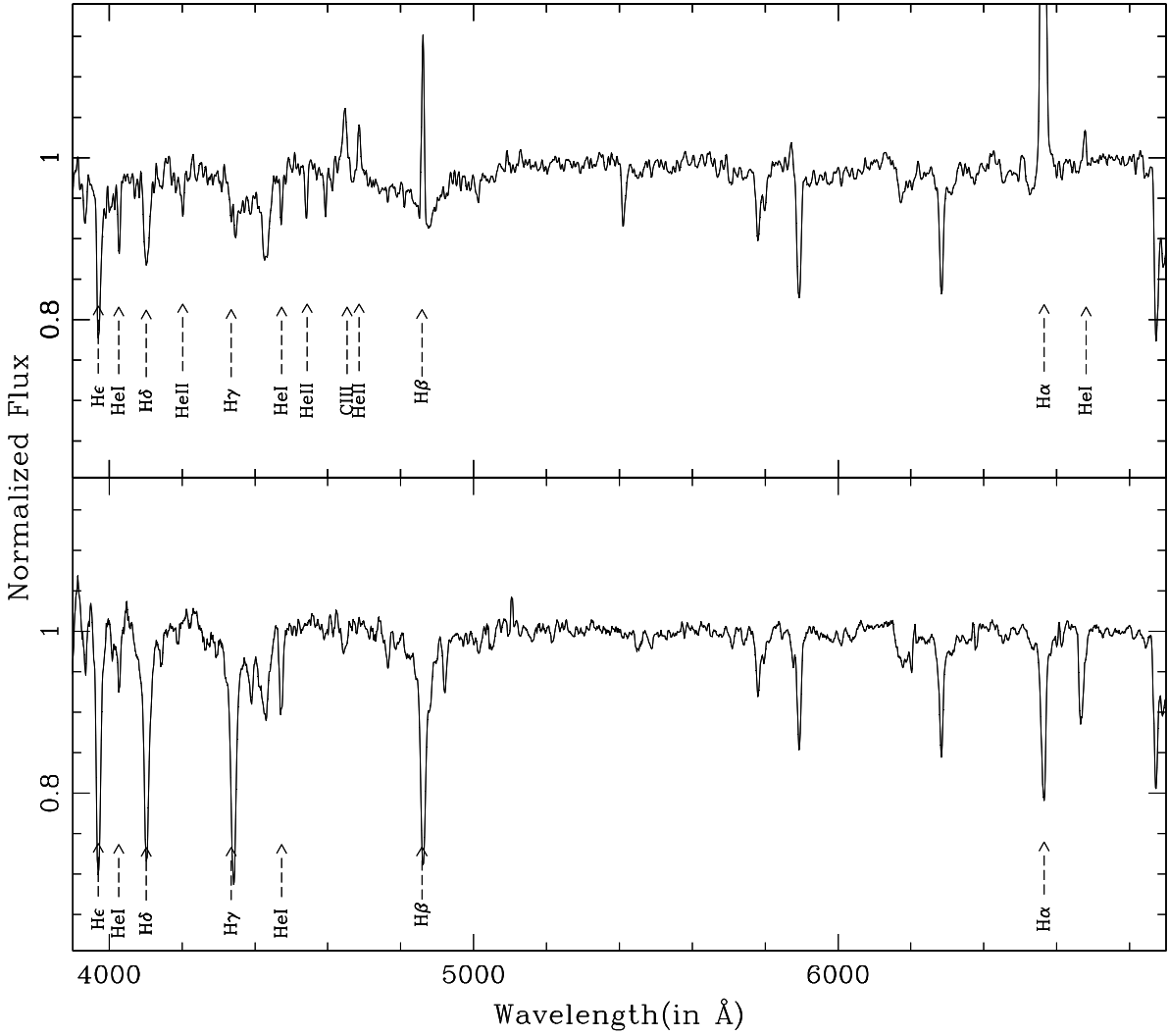}
\caption{Top: Flux calibrated normalized spectrum for the ionizing source M2.
Bottom: Wavelength calibrated normalized spectrum for the star M4. 
The lines identified for the spectral classification are marked in the figure. }
\label{spec}
\end{figure*}

\begin{figure*} \centering 
\includegraphics[scale = .5, trim = 10 20 10  80, clip]{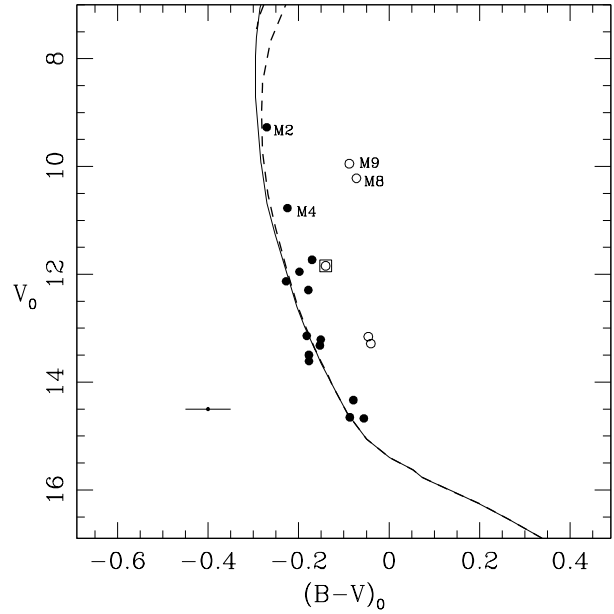}
\caption{$V_0/(B-V)_0$  CMD for stars lying within  $r \le 2^{\prime}$ of NGC 1624 
and having spectral type earlier than A0.  The filled and open circles are  the 
probable cluster  members and field stars, respectively.  The isochrones of  age  
2 Myr (solid curve)  and 4 Myr (dashed curve) by Girardi et al. (2002) corrected for the cluster distance are
also shown.  The labeled sources, numbered according to Deharveng et al. (2008), 
are further classified using low resolution spectroscopy to be of spectral 
class $\fiii-OV$ (see Sect. \ref{slitspec}).  The star shown by open square with open circle occupies 
a location near  to M8 and M9 stars in the $(J-H)/(H-K)$ colour-colour diagram and hence this
star could be a field giant.  The average error in the colour term is
given at the lower left side of the figure. }
\label{q}
\end{figure*}

\begin{figure*}
\centering 
\includegraphics[scale = .62, trim = 10 10 10 160, clip]{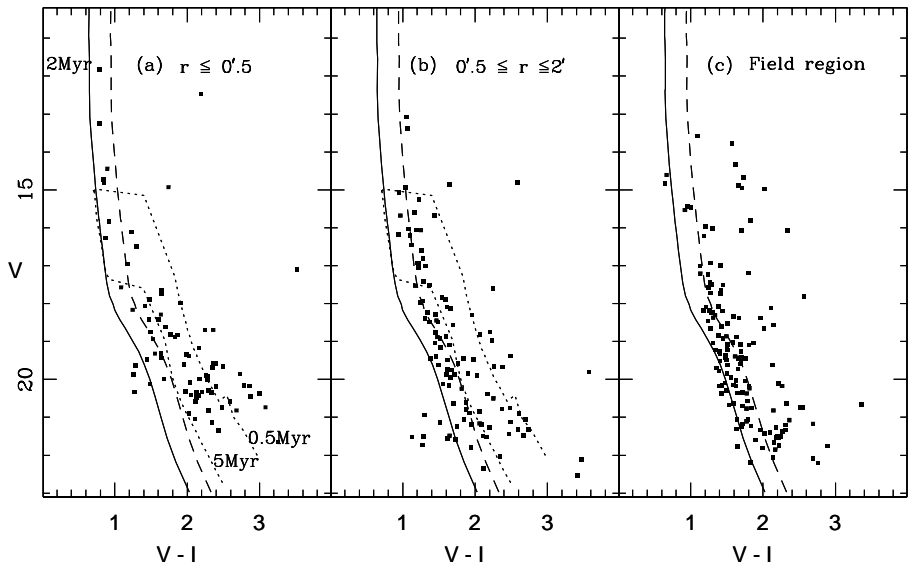}
\caption{ $V/(V-I)$ CMD for the stars  within (a): $r \le 0^\prime$.5 of  NGC 1624 
(b): within  $0^\prime.5 \le r \le  2^\prime$ of NGC 1624 (c): for stars in the control field. 
The  continuous curve is the  isochrone of 2 Myr
from Girardi et al. (2002) corrected for the cluster distance and
reddening  $E(B-V)_{min}$ = 0.76 mag, whereas the dashed curve is shifted 
for a reddening $E(B-V)_{max}$ = 1.0 mag.  The dotted curves are the PMS isochrone
for 0.5 and 5 Myr (Siess et al. 2000) shifted for the cluster distance and 
reddening  $E(B-V)_{min}$ = 0.76 mag} 
\label{cmd}
\end{figure*}

\begin{figure*} 
\centering
\includegraphics[scale = .5, trim = 10 120 50 150,clip]{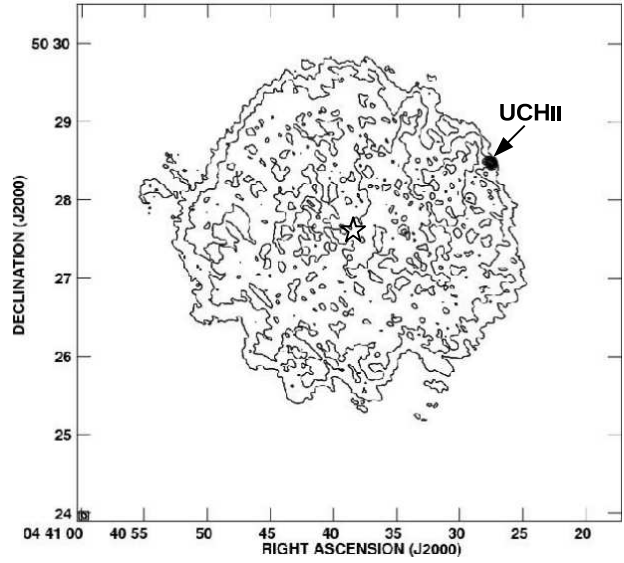}
\caption{GMRT high resolution map at 1280 MHz of Sh2-212 with a resolution of 
$\sim$ 4$^{\prime\prime}$.9 $\times$ 3$^{\prime\prime}$.2.  The contour levels are 
at 3, 4, 6, 9, 13, 18, 24  and  31 times of the rms noise 0.224 mJy/beam.  The 
star symbol represents the location of the cluster centre.} 
\label{1280}
\end{figure*}

\begin{figure*} 
\centering
\includegraphics[scale = .45, angle=-90, trim = 0 0 0 0, clip]{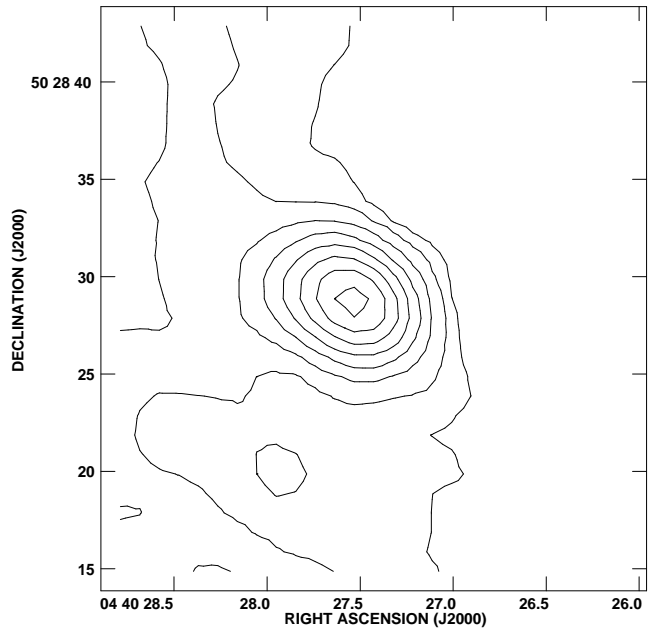}

\caption{ Enlarged map of UC\hii region at 1280 MHz from Fig. \ref{1280}. The 
contours are plotted above three times of the rms noise.  }
\label{610}
\end{figure*}

NIR data are   very useful tools to study the nature of young stellar population  within the 
star forming regions (SFRs).  Discriminating young stars in clusters from field stars is difficult. 
Young stars with strong infrared (IR) excess from disks and envelopes  can be identified using the 
NIR and mid-IR (MIR) observations.  We used the CFHT deep NIR photometry to study the PMS contents 
and KLF of NGC 1624. The  CFHT $K$-band  mosaic  image centered on the UC\hii region covering an 
area of  $5^{\prime}.2 \times 5^{\prime}.2$ is shown in Fig. \ref{cfht} (right panel), where  the 
ionizing source  is marked with a  white circle.  A very rich cluster is  apparent around the ionizing
source. Since the centre of NGC 1624 is located towards the eastern edge of the CFHT frame, eastern 
half of the cluster is  covered partially.  The observations covered an area  $\sim$ 9.6 arcmin$^2$ of
NGC 1624 and is  shown using a partial circle in Fig. \ref{cfht}. A region covering  an area $\sim$ 
3.1 arcmin$^2$  towards north of the cluster shown by a box in Fig. \ref{cfht}, is considered as 
the control field.   In the following sections, we discuss the NIR  CC  diagram and CMDs.

\subsection {Colour-Colour Diagrams}
\label{nircc}

\begin{figure*} \centering 
\includegraphics[scale = .5, trim = 0 10 0 50, clip]{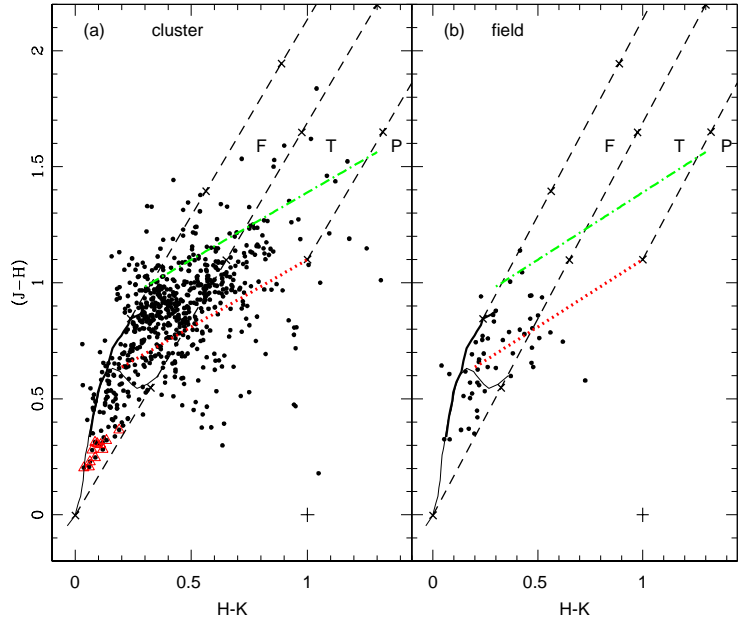}

\caption{$(J - H)/(H -  K)$ CC diagrams of sources detected in the  $J H K$ bands  
in (a)  NGC 1624  within  $\sim$ 9.6 arcmin$^2$ area  (b) control field of area 
$\sim$ 3.1 arcmin$^2$.  The locus for dwarfs  (thin solid curve) and giants 
(thick  solid curve) are  from Bessell  $\&$ Brett  (1988). The dotted  and 
dotted-dashed lines (red and green, respectively  in the online version) represent  the   
unreddened  and reddened ($A_V$ = 4.0 mag) locus of CTTSs (Meyer  et al.  1997).  
Dashed straight  lines represent  the reddening   vectors (Cohen et al. 1981). 
The crosses on the dashed lines are separated by $A_V$  = 5  mag.    The plots 
are classified in to three regions, `F', `T' and `P'.  The sources located in the `F' 
region are likely to  be the  reddened field stars, WTTSs or CTTSs with little or no 
NIR excess. The sources in the `T' region are considered to be candidate CTTSs with NIR 
excess and sources in the `P' region are the candidate Class I objects (see text for details).
 The sources marked using red triangles are the MS members identified using Q method 
(see Sect. \ref{reddening}).
The average photometric errors are shown in the lower right of each panel. }

\label{jhhk}
\end{figure*}

NIR and MIR photometry are useful tools to investigate the fraction of YSOs in a 
SFR. In the absence of ground based $L$-band observations or {\it Spitzer} based MIR 
observations, we used  $(J-H)$/$(H-K)$  CC diagram to identify the young 
stellar population in NGC 1624 (Hunter et al. 1995; Haisch et al. 2000; 2001; 
Sugitani  et al. 2002; Devine et al. 2008; Chavarr\'{i}a et al. 2010). The $(J-H)$/$(H-K)$  
CC diagrams  for  the cluster region (area $\sim$ 9.6 arcmin$^2$)  and 
the control  field  (area $\sim$ 3.1 arcmin$^2$)   are  shown in   Fig. \ref{jhhk}.  
The  thin and thick solid curves are the locations of unreddened MS and giant stars 
(Bessell $\&$ Brett 1988), respectively. The dotted  and dotted-dashed lines represent 
the locus of unreddened and reddened ($A_V$ = 4.0 mag) classical T Tauri stars (CTTSs; 
Meyer et al. 1997). The two long parallel  dashed lines are the reddening vectors for 
the early MS and giant type  stars (drawn from the base and tip of the two branches).  
One more reddening vector is plotted from the tip of the unreddened CTTS locus. The crosses on the 
reddening vectors are separated by  an $A_{V}$ value of  5 mag. The extinction ratios, 
$A_J/A_V = 0.265, A_H/A_V = 0.155$ and $A_K/A_V=0.090$, are adopted from  Cohen et al. (1981). 
The magnitudes, colours and the curves are  in CIT system. 

Presently YSOs are classified  as an evolutionary sequence spanning a few million years as:  
Class 0/Class I - the youngest embedded protostars surrounded by infalling envelopes and
growing accretion  disks; Class II - PMS stars with less active accretion disks and Class III - 
PMS stars with no disks or optically thin remnant disk (Adams et al. 1987). Following Ojha et
al. (2004a), we classified sources according to their locations in $(J-H)/(H-K)$ CC diagrams. 
The `F' sources are those located between  the reddening vectors projected from the intrinsic
colours of MS and giant stars. These sources are  reddened field stars (MS and  giants) or Class 
III/Class II sources with little or no  NIR excess (viz., weak-lined T Tauri sources (WTTSs)  
but some CTTSs may also be included). The sources located redward of region `F' are considered 
to have NIR excess. Among these, the `T' sources are located redward of `F' but blueward of the 
reddening line projected from the red end of the CTTS locus. These sources are considered to be 
mostly CTTSs (Class II objects)  with large NIR excesses (Lada \& Adams 1992). There may be an 
overlap in NIR colours of Herbig Ae/Be stars and T Tauri stars in the `T' region (Hillenbrand et 
al. 1992). The `P' sources are those located in the region redward of region `T' and are most 
likely Class I objects (protostellar-like) showing large amount of NIR excess. Here it is worthwhile 
to mention that Robitaille et al. (2006) have shown that there is a significant overlap between 
protostellar-like objects and CTTSs in the CC diagram.  

A comparison of the colour distribution of the sources in the cluster and control field (Fig. \ref{jhhk})
suggests that there is an appreciable difference between them. Significant fraction of sources  in the 
cluster region are concentrated between the unreddened and reddened  CTTS locus, whereas majority of  
sources in  the control field are mainly concentrated  in the `F' region.  Statistically, we can safely 
assume that  majority of  sources of the cluster region located between the unreddened and reddened 
CTTS locus are most likely to be  cluster members.  The comparison  also indicates that  the sources 
located in the `F' region   could be the  reddened field  stars but a  majority of them are likely 
candidate  WTTSs or CTTSs with little or no  NIR excess. The sources  lying towards the right  side of 
the reddening vector at the boundary of    `F' and `T' regions and above the unreddened CTTS locus 
can  be safely considered as YSO/NIR excess sources.  A total of 120 such sources have been detected 
within a $5^{\prime}.2 \times 5^{\prime}.2$ region which fall in the `T' region and above the unreddened CTTS 
locus.  However,  this number is certainly a  lower limit for the population of YSOs, as several of 
the cluster members   detected in the $H$ and $K$ bands   have not   been detected in the $J$-band.
Moreover, $L$-band or MIR observations would further increase 
the detection  of YSOs in the region. Hence the present $JHK$ photometry provides only a lower limit 
to the population of YSOs in NGC 1624. The distribution of YSOs in Fig. \ref{jhhk}  manifests 
that majority of them have $A_V$ $\le$ 4 mag. Some of the sources in `F' and `T' regions, which 
might be the candidate WTTSs/CTTSs, show $A_V$ values higher than 4 mag.  The $A_V$ for each star lying 
in `T' region has been estimated by tracing back to the intrinsic CTTS locus along the reddening vector. 
The $A_V$ for stars within the cluster region  (area $\sim$ 9.6 arcmin$^2$) and located in  the 
`F' region is estimated by tracing them back to the extension of the intrinsic CTTS locus 
(see Ogura et al. 2007; Chauhan et al. 2009 for details). The $A_V$ values thus calculated  for the 
sources in `F' and `T' regions are given in Table \ref{optdata}.   Twenty one  sources are found to have 
$A_V$ $\ge$ 6.0 mag, indicating that significant number of  cluster members in the region may still be 
embedded.

\subsection{The  colour-magnitude diagram}

\begin{figure*} \centering 
\includegraphics[scale = .7, trim = 10 20 10 120,  clip]{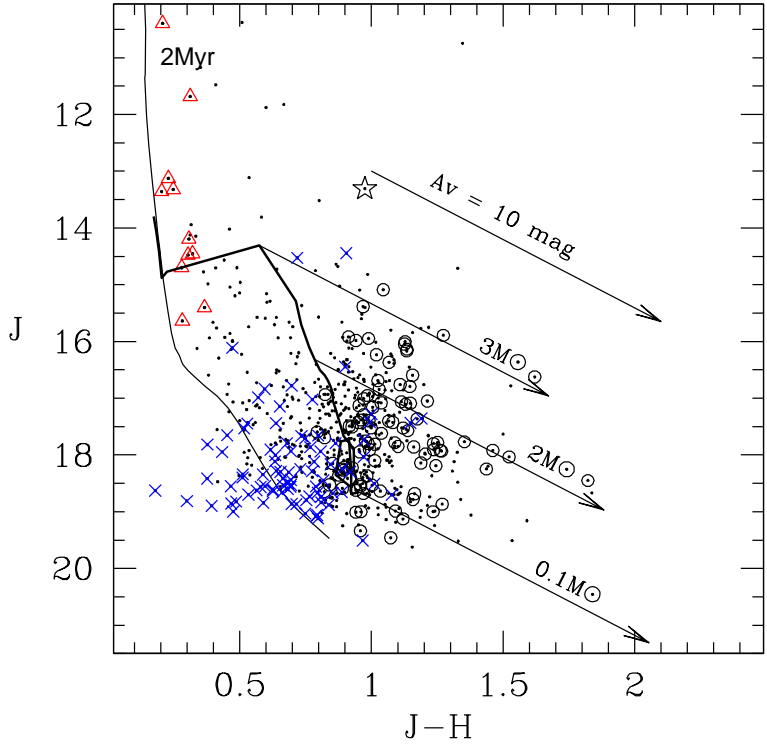}

\caption{$J/(J - H)$ CMD  for the sources within  $\sim$ 9.6 arcmin$^2$ area 
of NGC 1624. The encircled are the candidate NIR excess sources  and the crosses
are the sources which are lying below the CTTS locus.  The sources marked using red 
triangles are the MS members identified using Q method (see Sect. \ref{reddening}).
The star symbol represents candidate ionizing source of the UC\hii region. 
The thick solid curve represents the PMS isochrone of age 2
Myr  by Siess  et al. (2000) and the thin curve represents the  isochrone
of age 2 Myr by Girardi et al. (2002). Both the isochrones are corrected  for cluster distance and reddening. 
The continuous oblique  lines   denote  the reddening trajectories up to $A_V$ = 10 mag
for  PMS  stars  of  0.1,  2.0 and  3.0 $M_{\odot}$ for 2 Myr. }

\label{jhj}
\end{figure*}

Fig. \ref{jhj} shows  $J/(J-H)$  distribution of sources within $\sim$ 9.6
arcmin$^2$ area of NGC 1624. The encircled  are the NIR excess sources in this
region. The thick solid  curve denotes the locus of 2 Myr  PMS 
isochrone from Siess et al. (2000), which is the average age of NIR excess sources 
(see Sect. \ref{pms}, Fig.  \ref{yso}) and the thin curve is the  2 Myr  
isochrone from Girardi  et al. (2002). Both the isochrones are shifted for  the 
cluster distance and reddening. The continuous 
oblique  lines denote   the reddening trajectories up to $A_V$ = 10 mag for  PMS stars 
of 2 Myr age having masses 0.1, 2.0 and 3.0 $M_\odot$, respectively.  For the assumed age 
$\sim$ 2 Myr, reddening $A_V$ = 2.5 mag and distance =  6.0 kpc, the $J$-band detection 
limit of present observations corresponds to $M$ $\sim$ 0.1 $M_\odot$. In Fig. \ref{jhj}  
majority of NIR excess sources  ($\sim$ 98 \%) are seen to  have masses in the range 
0.1 to 3.0 $M_\odot$.

The CMD  indicates that the  stellar population in NGC 1624 
significantly comprises of low mass PMS stars similar to other SFRs
studied by Ojha et al. (2004a), Sharma et al. (2007),  Pandey et
al. (2008) and Jose et al. (2008). These results further support the
scenario  that the  high mass star forming regions are not devoid of
low mass stars (e.g., Lada \& Lada 1991; Zinnecker et
al. 1993; Tapia et al. 1997; Ojha et al. 2004a).  The distribution of 
stars located below the CTTS locus (cf. Fig. \ref{jhhk}) is shown by 
crosses in Fig. \ref{jhj}  which indicates that a  majority of these 
sources are likely to be  field stars.

The brightest  NIR excess source marked as a star symbol in Fig. \ref{jhj} 
is the candidate ionizing source  of the UC\hii region. The extinction to this star 
is estimated by tracing it back to the ZAMS along the reddening vector and found to be  $A_V$ $\sim$  
10.6 mag.  This extinction should be considered as an upper limit, as the star
shows NIR excess, therefore, $J$ and $H$ magnitudes might have been affected by the NIR  
excess emission. The photometric spectral 
type of this star comes out to be $\sim$ B0 which is in agreement with the spectral
type estimation based on our radio continuum observations (see Sect. \ref{ionized gas}).

\section{ Field star decontamination}
\label{field}

 Distinguishing cluster members from field stars is a significant 
challenge for photometric surveys of  clusters. 
To study the LF/MF, it is necessary to remove field star contamination
from the cluster region. Membership determination  is also crucial for 
assessing the presence of PMS stars because  both 
PMS and  dwarf foreground  stars occupy  similar positions above the ZAMS 
in  the CMDs.  As discussed in Sect. \ref{nir}, some of the YSOs can be
identified with the help of NIR excess, however this is not true for the 
diskless YSOs. An alternative is to study the
statistical distribution of stars in the cluster and field regions. 
Because  proper motion  studies are not available for the 
stars in the cluster region, we  used  following   statistical criteria  
to estimate the number of probable members of NGC 1624.

\begin{figure*} \centering 
\includegraphics[scale = .7, trim = 10 10 10 10, clip]{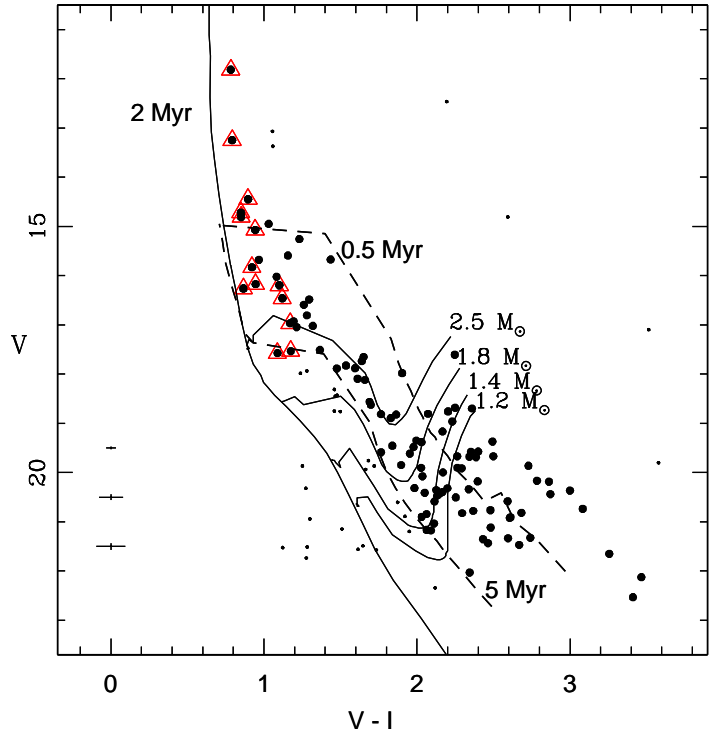}
\caption{Statistically cleaned $V/(V -  I)$ CMD (SCMD) for stars lying within
$r \le  2^{\prime}$ of NGC 1624.   The stars having PMS age $\le$
5 Myr are considered as representing the statistics of  PMS stars in 
the region and  are shown by filled circles. The sources marked using red 
triangles are the MS members identified using Q method (see Sect. \ref{reddening}). 
The  isochrone for 2 Myr 
age  by Girardi et al. (2002) and PMS isochrones of 0.5, 5 Myr  along with 
evolutionary tracks of different mass stars  by Siess et al.  (2000) are also
shown. All  the isochrones and tracks are corrected for  the cluster distance 
(6.0 kpc) and reddening ($E(B-V)$ = 0.76 mag).  The corresponding values of  
masses in solar mass  are given at the  right side of each track. Points shown 
by small dots are considered as non-members.  Average photometric errors in 
magnitude and  colour for different magnitude ranges  are shown  in the left 
side of the figure. }
 \label{calone}
\end{figure*}

To remove contamination  due to field  stars from the  MS and PMS  sample, we
statistically subtracted the contribution  of field stars from the
observed CMD of the  cluster region  using  the  following  procedure.
For any star in  the $V/(V-I)$ CMD of the  control field (Fig. \ref{cmd}c), the nearest
star in the cluster's  $V/(V-I)$   CMD (Figs. \ref{cmd}a and b) within   $V$ $\pm$ 0.125  and
$(V-I)$ $ \pm$ 0.065   was removed.   The  statistically cleaned $V/(V-I)$ CMD (SCMD) of the cluster 
region  is shown in Fig. \ref{calone}, which clearly shows a sequence towards red side of
the MS.  PMS  isochrones  by Siess et al. (2000) for  ages  0.5 and 5  Myr (dashed lines) and
2 Myr isochrone   by Girardi et al. (2002) (continuous  line)  are shown in  Fig. \ref{calone}.   
The evolutionary tracks by Siess et al. (2000)  for different  masses are also shown 
which are used to determine the  masses of  PMS cluster members.  Here we would like to remind 
the readers that the points shown by filled circles in Fig. \ref{calone} may not represent the actual 
members of the clusters. However, the filled circles should represent  the statistics of PMS stars in 
the region and the statistics has been used to study the MF of the cluster region 
(cf. Sect. \ref{imf}).

We followed the above technique for the field star decontamination of the NIR data as well. 
Since the area of the selected field region is smaller in comparison to the cluster region, we 
subdivided the cluster region in to three sub regions having area equal to the field region. The field star 
contamination from $J/ (J-H)$ CMD of the cluster sub regions was subtracted using the   $J/ (J-H)$ CMD 
of the field region in a similar manner as in the case of $V/(V-I)$ CMD.

\subsection{Young stellar population in NGC 1624}
\label{pms}

\begin{figure*} \centering
\includegraphics[scale = .7, trim = 10 20 10 120, clip]{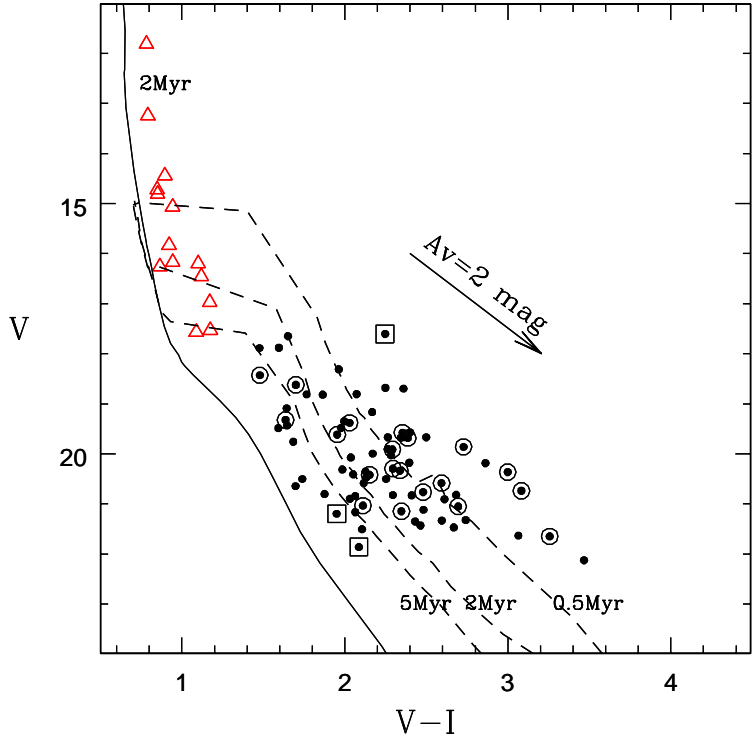}
\caption{$V/(V-I)$ CMD for the sources in NGC 1624 (area $\sim$ 9.6 arcmin$^2$) and
lying above the  unreddened  CTTS locus of the NIR  CC diagram (see Fig. \ref{jhhk}). The
encircled are the NIR excess  sources.  The sources marked using red triangles are the MS 
members identified using Q method (see Sect. \ref{reddening})  and those sources shown in box
are probable field stars.    Isochrone  for 2 Myr age (solid curve)  
by Girardi et al. (2002) and PMS isochrones of age   0.5, 2 and  5 Myr (dashed curves) by Siess et
al. (2000) are also shown. All  the isochrones are corrected for  the cluster  distance and 
reddening. The arrow indicates  the reddening vector for $A_V$ = 2 mag. }
\label{yso}
\end{figure*}

 It is found that nineteen percent of the candidate PMS stars located above the intrinsic CTTS locus
(cf.  Fig. \ref{jhhk})  have optical counterparts in $V$-band within  9.6 arcmin$^2$ area.
The $V/(V-I)$ CMD for these sources is shown in Fig. \ref{yso}. The encircled are the NIR 
excess sources which are the likely candidate YSOs (see Sect. \ref{nircc}). PMS isochrones by 
Siess et al. (2000) for 0.5, 2, 5 Myr (dashed curves) and   isochrone for 2 Myr by Girardi et al. (2002; 
continuous curve) corrected for cluster distance and reddening are also shown.  Fig. \ref{yso} reveals 
that  majority of the  sources have  ages $\le$  5 Myr   with a possible age spread of $\sim 0.5 - 5$ Myr 
and $\sim$ 75$\%$ of the NIR excess sources show ages $\le$ 2 Myr. Since the reddening vector in  $V/(V-I)$ 
CMD (see Fig. \ref{yso}) is nearly parallel to the PMS isochrone, the presence of variable extinction in 
the region will not affect the age estimation significantly. Therefore the age spread indicates a possible 
non-coeval star formation in this region. 

The membership of the YSOs shown in Fig. \ref{yso} is calculated using the following  procedure. Each YSO
is corrected for its reddening calculated in the Sect. \ref{nircc}.  The intrinsic $(V-I)$ colour thus 
obtained is then compared with the  PMS isochrones of varying ages from 5 Myr to 0.1 Myr. 
The $M_V$ value of each YSO is obtained from the best matching isochrone  and hence the distance modulus. The 
sources lying within $3\sigma$ of the distance modulus obtained in Sect. \ref{distance} are considered as 
the probable cluster members. It is found that three sources do not satisfy the above criteria and has been 
considered as non-members. These three sources are  marked using box in  Fig. \ref{yso}.

A  comparison of  Fig. \ref{yso} with the field star decontaminated CMD  shown in Fig. \ref{calone} 
reveals  a nice resemblance,  suggesting that  the statistics of  
PMS sources selected on the basis of  SCMD can be used  to study the IMF 
of PMS population  of NGC 1624. As most of the sources in Fig. \ref{yso} are located in the 
PMS region,  it can be safely assumed that the sources lying above the unreddened  CTTS
locus of Fig. \ref{jhhk} are likely  cluster members.  Thus sources falling in the `F' region 
(see Fig. \ref{jhhk}) are likely to be  WTTSs or CTTSs with little or no NIR excess and those in 
the `T'  region are the candidate CTTSs with NIR excess. However, Fig. \ref{yso} does not 
show any trend in age distribution  between  these sources.  A comparison of Figs.  
\ref{jhj} and \ref{calone} confirms  that most of the YSOs have masses $\le$  3.0  $M_\odot$.

The fraction of NIR excess sources in a cluster is also  an  age indicator because the 
disks/envelopes become optically thin with age (Haisch et al. 2001;  Carpenter et al. 2006; 
Hern\'{a}ndez et al. 2007). For young embedded clusters having  age $\le$ $ 1 \times 10^6$ yr, 
the disk fraction obtained from $JHK$  photometry is  $\sim$ 50\% (Lada et al. 2000; Haisch et al.  
2000). Whereas the fraction reduces to $\sim$ 20\%  for the clusters with age $\sim$ 2 -
3 $\times$ $10^6$ yr  ( Lada \& Lada 1995; Haisch et al. 2001; Teixeira et al. 2004; Oliveira 
et al. 2005). After correcting for the field star contamination and photometric incompleteness, 
the fraction of NIR excess sources in  an area $\sim$ 9.6 arcmin$^2$ of  NGC 1624 is estimated to be
$\sim$ 20\%.  There are 31 $J$ drop-out sources falling within our error criteria. Based on the 
colour and spatial  distribution of these $J$ drop-out sources (see Sect. \ref{distribution}), we presume 
that they can be included in the list of candidate YSOs and hence the NIR excess fraction increases to
$\sim$ 25\%.  This suggests an age of $\sim$ 2 - 3 $\times$ $10^6$ yr for this cluster which is 
in agreement with the age estimation derived using the PMS  evolutionary tracks in the optical CMD
(cf. Fig. \ref{yso}). This NIR excess fraction is to be considered as a lower limit  to the actual 
YSO fraction of the cluster  as we do not have  $L$-band observations for this cluster.
However, Yasui et al. (2009) point out that the  disk fraction from only $JHK$ data are about 0.6 
of those from $JHKL$ data and the lifetime estimation from $JHK$ data is
basically identical to that from $JHKL$ data.  Therefore, despite a little larger uncertainty, 
the disk fraction from  $JHK$ data alone  should still be effective even without $L$-band data.
Here it is worthwhile to point out that  in the case of Cep OB3B, Getman et al. (2009) have shown 
that the disk frequency  depends on the distance from the exciting stars, as  massive stars 
can photo-evaporate the disk around young stars. Also, Carpenter et al. (2006) have found 
evidence for mass dependent circumstellar disk evolution in the sense that the mechanism for disk
dispersal operates  less efficiently for low mass stars.  Hence, keeping in mind the uncertainties  
mentioned above, the age estimation based on the disk frequency must be considered as an approximate 
estimation. 

In order to check if there is any mass dependence of the NIR excess fraction, we divided 
the optically identified PMS members (shown in Fig. \ref{yso}) in to three mass bins i.e., 
2.5 - 1.5 $M_{\odot}$, 1.5 - 1.0 $M_{\odot}$ and  1.0 - 0.6 $M_{\odot}$ using the evolutionary 
tracks by Siess et al. (2000). After applying  the completeness correction in each magnitude bin,  
we obtained the NIR excess fraction as 23\%, 24\% and 37\%,  respectively for the above mentioned 
mass bins. Hence, there is an evidence of mass dependent evolution of circumstellar  disk as 
explained by Carpenter et al. (2006). However, this estimation has to be considered as a lower 
limit, as only 19\% of the identified NIR PMS stars have the optical counterparts.

Deharveng et al. (2005; 2008) have identified signs of  recent star formation in  Sh2-212. They 
estimated the  age of the massive star associated with the UC\hii region located at the periphery of 
Sh2-212 as $\sim$ 0.14 Myr  on the basis of dynamical size of the UC\hii region.  This indicates 
that the UC\hii region is relatively  young as  compared to  the YSOs within the cluster region. 
The bright rim feature at one end of the UC\hii region (see Fig. 2 of Deharveng et al. 2008)  
also suggests   that the UC\hii region might have formed at a later  evolutionary stage  of the \hii 
region as a second generation object. 

\section {Spatial distribution of YSOs}
\label{distribution}

\begin{figure*} \centering 
\includegraphics[scale = .8, trim = 10 30 10 170, clip]{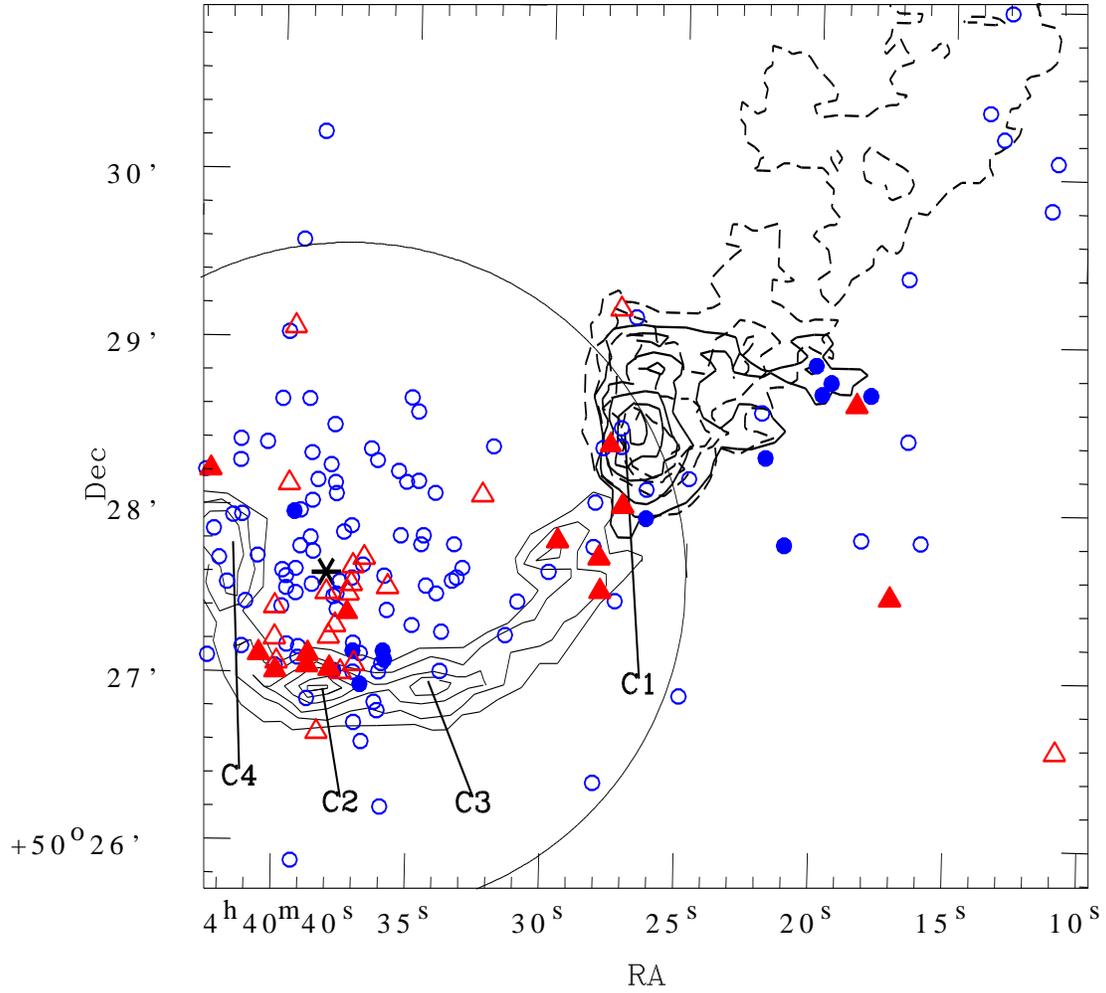}
\caption{ Spatial distribution of YSOs (blue circles in the online version) 
and the $J$ drop out sources (red triangles).  The sources with $H-K \ge 1.0 $ mag are shown  
using  filled circles and triangles,  respectively  and the asterisk 
represents the  centre of NGC 1624. The contours represent the $^{13}$CO(2-1)
emission map from Deharveng et al. (2008) in the velocity range between 
-34.0 km s$^{-1}$ to -32.7 km s$^{-1}$ (continuous thin contours),   -36.1 $km s^{-1}$ 
to -35.1 $km s^{-1}$ (continuous thick contours)
and -36.8 km s$^{-1}$ to -35.9 km s$^{-1}$ (dashed contours), respectively. The partial circle represents the
2$^\prime$ boundary of the cluster.}

\label{co}
\end{figure*}

Fig. \ref{co} displays the spatial distribution of YSOs (blue circles;  likely Class II sources)  
identified on the basis of NIR excess characteristics (cf. Fig. \ref{jhhk}) along with the CO emission 
contour map from Deharveng et al. (2008) for four condensations and filament. The $J$ drop-out sources are 
shown using red triangles. The molecular condensations make a semi-circular ring towards the  southern side of 
Sh2-212. Fig. \ref{co} reveals that  majority of YSOs are located close to the cluster centre within a 
radius of 0.5 arcmin (i.e., within the cluster core radius of $\sim$ 0.9 pc; cf. Sect.  \ref{rd}), however,
several other YSOs are found to be distributed outside of this radius along the thin semi-circular ring 
and filamentary structure. Interestingly, there is an apparent concentration of YSOs  just at the boundary 
of the clump C2.

In Fig. 15 we have shown the  $ K/(H-K) $  CMD  for  all the sources detected in this region. The encircled 
are the YSOs and the red triangles are the $J$ drop-out sources. It is evident from the CMD  that majority 
of  YSOs have $(H-K)$ colour in the  range  $\sim$ 0.6 - 0.8 mag. However a  significant number of sources  
appear to be redder ($H-K$ $ \ge 1.0$  mag).  The spatial distribution of sources having $(H-K) \ge 1.0$ mag 
has been shown  in Fig. \ref{co} with filled circles (i.e., YSOs) and triangles ($J$ drop-out sources), 
respectively and this figure reveals  
a higher density  of reddened sources near the clump C2. The larger  value of $(H-K)$  ($\ge$ 1.0  mag)
could be either due to higher extinction, as most of these sources are lying within/very close  to the CO 
distribution, or could be their intrinsic colour due to large NIR excess.

If the origin of this colour excess is merely from the interstellar extinction, then one must expect an 
increment in the value of $A_V$  by $\sim$ 12 mag as compared to the sources located close to the cluster center. 
In order to investigate the spatial distribution of extinction in the region, we plot radial variation of  $A_V$ 
in Fig. \ref{radial} (left panel). It is evident from the Fig. \ref{radial} that $A_V$  is almost constant 
within an 80 arcsec cluster radius. Hence, we can  presume that the origin of colour excess could be intrinsic 
in nature. This fact indicates an age sequence in the sense that YSOs located/projected over the  semi-circular 
ring of molecular condensations are younger than those lying within the core of the cluster.

To further elucidate the youth of the YSOs  located/projected over the  semi-circular ring  of molecular 
condensations, we plot  radial variation of NIR excess,  $\Delta(H-K)$, defined as the horizontal displacement 
from the reddening vector at the boundary of ‘F’ and ‘T’ regions (see Fig. \ref{jhhk}). NIR excess is considered
to be a function of age.  An enhancement in the mean value of  $\Delta(H-K)$ at $\sim$ 45 arcsec, i.e.,
near the periphery of the semi-circular ring is 
apparent in Fig. \ref{radial} (middle panel).  In the right panel we plot the radial variation of $(H - K)$ 
colour  of YSOs and $J$ drop-out sources using dashed and solid histogram, respectively. 
The enhancement in the mean $(H - K)$  value at the same location is apparent in this figure as well.
However, we have to keep in mind  the possibility of photo-evaporation of the disk around YSOs  lying within 
the core of the cluster due to stellar radiation of massive star at the centre of the cluster.

The above facts indicate that the sources near the molecular material are intrinsically redder and 
support the scenario of possible sequential star formation towards the direction of 
molecular clumps.  It is interesting to mention that the distribution of YSOs in the NGC 1624 region is 
rather similar to the distribution of Class II sources in other star forming regions. e.g., RCW 82 
(Pomar\`{e}s et al. 2009), RCW 120 (Zavagno et al. 2007) and Sh2-284 (Puga et al. 2009). Majority of  
Class I sources in the case of RCW 82 and RCW 120 are found to be associated with the molecular material 
at their periphery and none are found around the ionizing source. The association of Class I sources with 
the molecular material manifests the recent star formation at their periphery. If star formation in  Sh2-212 
region is similar in nature to RCW 82 and RCW 120, one would expect a significant number of Class I sources 
in the surrounding molecular material. Unfortunately, the absence of MIR observations hampers the detailed study 
of the probable young sources lying towards the collected molecular material. However, the YSOs having 
$(H-K) \ge 1.0$ mag, which are expected to be the youngest sources of the region, are found to be distributed 
around the molecular clumps detected by Deharveng et al. (2008).  It is interesting to mention that in the 
case of RCW 82, the YSOs having  $(H-K) \ge 1.0$ mag are found to be associated with the molecular emission 
surrounding the \hii region. Many of these sources are not observed in the direction of  molecular emission 
peaks, but are located on the borders of the condensations (Pomar\`{e}s et al. 2009). A similar distribution of 
YSOs (having $H-K \ge 1.0$) can be seen  in the present study at the border of the clump C2.

According to Deharveng et al. (2008), the massive YSO associated with the UCH II region (clump C1)
might have formed as a result of the collect and collapse process due to the expansion of the \hii region. If 
the sources lying towards the molecular clump C2 and along the filament are formed as a result of the collect 
and collapse process, these sources must be younger than the ionization source by about 2 - 3 Myr as the model 
calculation by Deharveng at al. (2008) predicts the fragmentation of the collected layer after 2.2 - 2.8 Myr 
of the formation of the massive star in Sh2-212. Since the ionization source is an  O6.5 $\pm$ 0.5 MS star, the 
maximum age of the ionization source should be of the order of its MS life time, i.e., $\sim$ 4.4 Myr 
(cf. Sect. \ref{distance}).   On the basis of the  present analysis we can indicate that the sources with   
$(H-K) \ge 1.0$ seem to have a  correlation with the  semi-circular ring  of molecular condensations and 
should be younger than the age of the ionization source of the region. However in the absence of optical 
photometry, the reliable age estimation of these YSOs is not possible. Since the distribution of youngest 
YSOs on the border of clump C2 has a resemblance to the distribution of   Class I/ II YSOs in RCW 82, 
the formation of these YSOs could be due to  the result of small-scale Jeans gravitational instabilities in the 
collected layer, or interactions of the ionization front with the pre-existing condensations as suggested 
by Pomar\`{e}s et al. (2009) cannot be ignored. 

\section{Initial Mass Function}
\label{imf} 
The distribution of stellar masses that form in a star formation event in a given volume of 
space is called IMF  and together with star formation rate, the IMF dictates the evolution 
and fate of galaxies and star clusters (Kroupa 2002). Young  clusters are important  tools to 
study  IMF since their MF can be considered  as IMF  as they  are too young  to loose significant 
number   of members  either  by   dynamical  or  stellar evolution.  To study  the IMF  of NGC 1624  
we  used  the data within $r \le 2^\prime$.

The MF is often expressed by the power law,  $N (\log m) \propto m^{\Gamma}$  and the slope of the 
MF is given as     

$$ \Gamma = d \log N (\log m)/d \log m $$

\noindent where $N  (\log m)$ is the  number of stars per  unit logarithmic mass interval.  For the 
mass range $0.4  < M/M_{\odot} \le 10$,  the classical   value derived  by Salpeter  (1955)  for the
slope of MF is, $\Gamma = -1.35$. 

 Since the NIR data is  deeper, we expect to have a better detection  of YSOs towards the fainter end 
in comparison to the optical data. Therefore we estimated  the IMF using the optical and NIR data independently.

\subsection {IMF from optical data}

With the  help of SCMD shown in Fig. \ref{calone}, we can derive the MF 
using  theoretical evolutionary models.  A mass$-$luminosity relation is needed to convert the
derived magnitude for each star to a mass. For  the MS stars (see Fig. \ref{q}),  
LF was converted to MF using the theoretical model by  Girardi et al. (2002) for 2 Myr
 (cf. Pandey et al. 2001; 2005).  The MF for PMS 
stars was  obtained by counting the number of stars in various mass bins (shown as  evolutionary 
tracks in Fig. \ref{calone}). Necessary corrections for data incompleteness were taken into 
account for each magnitude bin to calculate  the MF. The  MF of NGC 1624 is  plotted in
Fig. \ref{mf}.   The slope, $\Gamma$ of  the  MF    in  the mass range $1.2 \le M/M_{\odot}<27$  
can be  represented by a power law. The slope of the MF for the mass range  $1.2 \le M/M_{\odot}<27$ 
comes out to be,  $\Gamma$ = $-1.18\pm0.10$, which is slightly shallower than the Salpeter value 
(-1.35). We conclude that within an acceptable margin, the slope of IMF for the cluster NGC 1624 
is comparable to the Salpeter (1955) value.

\subsection {IMF from NIR data} 

We also estimate the IMF using J -band luminosity function (JLF). 
We preferred $J$-band over $K$-band as the former  is least affected by the NIR excess.  After
removing the field star contamination using  the statistical subtraction  as 
explained in Sect. \ref{field}, we applied the completeness correction to the $J$-band 
data.  Assuming an average age of 2 Myr for the PMS stars, distance
6.0 kpc and average reddening $A_V$ = 2.5 mag, the $J$ magnitudes were
converted  to mass using the 2 Myr PMS isochrone by Siess et
al. (2000). For MS stars, the  mass-luminosity relation is taken from
Girardi et al. (2002). Completeness  of the $J$-band data was $\sim$
90 \% at $J$ = 18  mag ($\sim$ 0.65 $M_\odot$). In Fig. \ref{mf_ir},
we have shown the MF derived for NGC 1624 (within the area of $\sim$
9.6 arcmin$^2$) in the mass range $0.65 \le M/M_{\odot}<27$. The
linear fit gives a slope  $\Gamma$ = $-1.31\pm0.15$ which is in
agreement with the  Salpeter (1955) value. The MF ($\Gamma$ =
$-1.18\pm0.10$) derived using optical data is slightly shallower than
that of IR data. However both the slopes are within error and can be
considered to be in agreement.   

Here we would like to point out that the estimation of IMF depends on
the models used. We are pursuing studies of  few young clusters, hence
a  comparative study of IMFs of various young clusters obtained using
similar techniques will give useful information about IMFs. 
Our recent studies on  young clusters (age $\sim 2 - 4 $  Myr), viz., 
NGC 1893 (Sharma et al. 2007), Be 59 (Pandey et al. 2008) and Stock 8 (Jose et al. 2008) 
have yielded the value of $\Gamma$ for stars more massive than $\sim $ 1 - 2 $M_\odot$ 
as -1.27 $\pm$ 0.08, -1.01 $\pm$ 0.11 and -1.38 $\pm$ 0.12, respectively. 
A comparison of the MF in the case of NGC 1624 and the clusters  mentioned above indicates that
the MF slope towards massive end (i.e., M $\ge 1 M_\odot$) in general, is comparable
to the Salpeter value (-1.35).

\section {K-band luminosity function}

The  KLF is frequently used in studies of
young clusters as a powerful tool to constrain its  age and IMF. Pioneering 
work on the interpretation of KLF was presented by Zinnecker
et al. (1993). During the last decade several  studies have been carried out with the aim of 
determining the KLF of young clusters  (e.g., Muench et al. 2000; Lada \&  Lada 2003;
Ojha et  al. 2004b; Sanchawala et al. 2007; Sharma et al. 2007; Pandey
et al. 2008; Jose et al. 2008). We have used CFHT $K$-band data to
study the KLF of NGC 1624. Because the CFHT observations   did not include
the entire cluster region, we restricted the KLF   study  to a region within
$\sim$ 9.6 arcmin$^2$ area of NGC 1624 (see  Sect. \ref{nir}).  

In order to convert the observed KLF  to the true KLF, it is necessary to correct the data 
incompleteness   and field star contamination. We applied the CF (see Sect. \ref{cfhtdata}) for 
the data incompleteness.  The control field having an  area $\sim$ 3.1 arcmin$^2$ shown in Fig. \ref{cfht}
has been used to remove the field star contribution. We applied a correction factor 
to take into account the different areas of cluster and control field regions.  
The field star population towards the direction of NGC 1624 is also estimated  by using the
Besan\c con Galactic  model of stellar population  synthesis (Robin
et al.  2003) using a similar procedure as described by Ojha et al. (2004b). 
The star counts were  predicted using the  Besan\c con
model towards the direction of the control field. An advantage of
using this model is that we can simulate foreground ($d<6.0$ kpc) and
background  ($d>6.0$   kpc)  field star populations
separately. The use of this model allows us to apply the  extra cloud
extinction to the background stars. The foreground population  was
simulated using the model with $A_V$ = 2.36 mag ($E(B-V) = 0.76$ mag;
ref. Sect. \ref{reddening}) and $d < 6.0$ kpc.   The background
population ($d>6.0$ kpc) was simulated with an extinction  value
$A_V$ = 4.0 mag (see Sect. \ref{nircc}).   Thus we  determined  the  fraction  of  the
contaminating  stars (foreground + background)  over  the  total
model counts. The scale factor  we obtained to the control field
direction was close to 1.0 in all the  magnitude bins. This indicates
that the moderate extinction of $A_V$ $\sim$ 4.0 mag is unlikely to have any
significant effect on the field star distribution  at this distance.
Hence, we proceeded our analysis of KLF with the field star counts
obtained from the observed control field.  The completeness corrected and  
field star  subtracted  KLF for NGC 1624 is shown  in Fig. \ref{klf}.  

The KLFs of young embedded clusters are known to follow power-law
shapes  (Lada et al. 1991; 1993) which is  expressed as:

\begin{center}
${{ \rm {d} N(K) } \over {\rm{d} K }} \propto 10^{\alpha K}$
\end{center}

where ${ \rm {d} N(K) } \over  {\rm{d} K }$ is the number of stars per
0.5 mag   bin  and   $\alpha$  is  the   slope  of   the  power
law. The KLF for NGC 1624  shown in  Fig. \ref{klf}  (solid line), yields
a slope   $0.30\pm0.06$ for the  range $K$ = 13.5 - 17.5 mag, which is slightly  lower than the average
value  of slopes  ($\alpha  \sim 0.4$) for  young clusters of similar ages (Lada  et
al. 1991; Lada \& Lada  1995; Lada \& Lada 2003).   However,  a break in the power 
law can be noticed at $K$ = 15.75 mag and the KLF seems to be flat in the magnitude range 15.75 - 17.5.   
The slope of the KLF in the magnitude range  13.5 - 15.75 (dahsed line in Fig. \ref{klf}) comes out to 
be 0.44 $\pm$  0.11 which is comparable  
with the average value of slopes for young clusters. A turn off in the KLF has also been observed 
in a few young clusters. e.g., at $K \sim$ 14.5 mag and $K \sim$ 16.0 mag in the case of Tr 14 
(distance $\sim$ 2.5 Kpc; Sanchawala et al. 2007) and NGC 7538 (distance $\sim$ 2.8 Kpc; Ojha et al. 2004), 
respectively.

KLF slope is an age indicator of  young clusters. For clusters up to 10 Myr 
old, the KLF slope  gets steeper as the cluster gets older  (Ali \& Depoy 1995; 
Lada \& Lada 1995). However, there is no precise  age - KLF relationship 
in the literature due to  huge uncertainty  in their correlation (Devine et al. 2008). 
There are many studies on KLF of young clusters.  The studies by
Blum et al. 2000; Figuer\^{e}do et al. 2002; Leistra et al. 2005; 2006; Devine at al. 2008 
indicate that the KLF slope  varies from 0.2 -0.4 for  clusters younger than 5 Myr. 
The KLF of NGC 1624 is worth comparing with the recent studies of
young clusters viz; NGC 1893 (Sharma et al. 2007), Be 59 (Pandey et al. 2008) and Stock 8
(Jose et al. 2008), since  all the KLFs are obtained using a similar technique. The slope
of the KLF ($\alpha = 0.30\pm0.06$) obtained for NGC 1624 in the magnitude range 13.5 - 17.5 is comparable with
those obtained for NGC 1893 ($\alpha = 0.34\pm0.07$), Stock 8 ($\alpha
= 0.31\pm0.02$) and Be 59  ($\alpha = 0.27\pm0.02$).

\section{Summary}  
We have carried out a comprehensive multi-wavelength study of the young cluster NGC 1624  associated
with the \hii region  Sh2-212. Sh2-212  is thought to have experienced
`Champagne flow'  and the molecular clumps along with the UC\hii
region at the  periphery are suggested as the possible outcome of the
collect and collapse phenomena.  In our present study, an attempt has
been made to determine the basic properties  of NGC 1624 as well as to
study the nature of stellar contents  in the region  using optical $UBVRI$
photometry, optical spectroscopy of four stars, radio continuum observations 
from GMRT  along with NIR $JHK$ archival data from 2MASS and CFHT. 

From optical observations of massive stars, reddening ($E(B-V)$) 
in the direction of NGC 1624 is found to vary between  0.76 to 1.00 mag 
and distance is estimated to be $6.0 \pm 0.8$ kpc.  The maximum post-main-sequence age of the
cluster is estimated as $\sim$  4 Myr.  Present spectroscopic 
analysis of the ionizing source indicates  a  spectral class of O6.5V. We 
used $JHK$ colour criteria to identify sources with  NIR excess and found 
120 candidate YSOs in the region.  Majority of the YSOs have $A_V \le$ 
4.0 mag and  masses  in the range $\sim$ 0.1 - 3.0  $M_\odot$. Distribution of these YSOs on the CMD indicates 
an age spread of  $\sim$ 0.5 - 5 Myr  with an average age of  $\sim$ 2-3 Myr, suggesting  
non-coeval star formation in NGC 1624. The lower limit for the  NIR 
excess fraction  on the basis of $JHK$ data is found to be $\sim$ 20\% which  
indicates an average age  $\sim$ 2 - 3 Myr for YSOs in NGC 1624. From the radio continuum 
flux, spectral type of the ionizing source of the UC\hii region is estimated to be
$\sim$ B0.5V. 

A significant number of YSOs are located close to the  cluster centre 
and a few YSOs are seen  to be located/projected over the molecular clumps detected by Deharveng et al. 
(2008), as well as farther  away from the clumps. We detect an enhanced density 
of reddened YSOs located/projected close to the molecular clump C2. The  NIR excess and $(H-K)$ colour
distribution of these sources show indication of an  age sequence in the sense that the YSOs 
located/projected near the clump C2 are younger than  those located within the cluster core.   

The slope of the MF, $ \Gamma$, derived from optical data, in  the
mass range $1.2 \le M/M_{\odot}<27$ can be represented by -1.18 $\pm$
0.10.  Whereas NIR data, in the mass range $0.65 \le M/M_{\odot}<27$
yields $ \Gamma$ = -1.31 $\pm$ 0.15. Thus MF fairly agrees with the
Salpeter  value (-1.35).   Slope of the KLF for NGC 1624  in the magnitude range 13.5 - 17.5 is found to
be 0.30 $\pm$ 0.06 which is smaller than the average value ($\sim$0.4)
obtained for young  clusters of similar ages (Lada et al. 1991; Lada
\& Lada 1995; Lada \& Lada 2003), however, agrees well with the values
0.27 $\pm$ 0.02  for Be 59 (Pandey et al. 2008); 0.34 $\pm$ 0.07 for
NGC 1893 (Sharma et al. 2007) and  0.31 $\pm$ 0.02 for Stock 8 (Jose
et al. 2008).  However, there is a clear  indication of break in 
the power law at $K$ =15.75 mag. The KLF slope in the magnitude range 
13.5 - 15.75  can be represented by $\alpha = 0.44 \pm 0.11$ and the KLF 
slope  is found to be flat in the magnitude range 15.75 - 17.5. 
 
\section{Acknowledgments}

Authors are thankful to the referee Dr. Antonio Delgado for his useful comments
which has improved contents and presentation of the paper significantly.
We thank the staff of  IAO, Hanle and its remote control station at CREST, 
Hosakote, ARIES, Naini Tal, and GMRT, Pune, India for  their assistance during  
observations. This publication  makes use of  data from  the Two  Micron All  
Sky Survey, which is  a joint project of  the University of  Massachusetts and 
the Infrared  Processing  and   Analysis  Center/California  Institute  of
Technology,   funded   by   the   National   Aeronautics   and   Space 
Administration and the National  Science Foundation.  This research used 
the facilities of the Canadian Astronomy Data Centre operated by  the National 
Research Council of Canada with the support of the Canadian Space Agency.  We 
thank Annie Robin for letting us use her model of stellar population synthesis.
JJ is thankful for the financial support for this study through a stipend from the DST 
and CSIR, India.  

\section{REFERENCES}

Adams F. C., Lada C. J., Shu F. H. 1987, ApJ, 312, 788\\ 
Ali, B.,  Depoy, D. L., 1995, AJ, 109, 709\\
Bessell, M.,  Brett, J. M., 1988, PASP, 100, 1134\\
Blitz, L., Fich, M., Stark, A. A., 1982, ApJS, 49, 183\\
Blum, R. D., Conti, P. S.,  Damineli, A., 2000, AJ, 119, 1860\\
Caplan, J., Deharveng, L., Peña, M., Costero, R.,  Blondel, C., 2000, MNRAS, 311, 317\\
Carpenter, J. M., 2001, AJ, 121, 2851\\
Carpenter, J. M., Mamajek, E. E., Hillenbrand, L. A.,  Meyer, M. R. 2006, ApJ, 651, L49\\
Chabrier, G., 2005, The Initial Mass Function 50 Years Later, 327,  41\\
Chavarr\'{i}a, L., Mardones, D., Garay, G., Escala, A., Bronfman, L.,  Lizano, S., 2010, ApJ, 710, 583\\
Chauhan, N., Pandey, A. K., Ogura, K., Ojha, D. K., Bhatt, B. C., Ghosh, S. K., Rawat, P. S., 2009, MNRAS,396,964\\
Chini, R.,  Wink, J.E., 1984, A\&A, 139, L5\\
Chini, R.,  Wargau, W. F., 1990, A\&A, 227, 213\\ 
Cohen, J. G., Persson, S. E., Elias, J. H.,  Frogel, J. A., 1981, ApJ, 249, 481\\
Conti, P.S.,  Alschuler, W. R., 1971, ApJ, 170, 325\\       
Cutri, R. M., Skrutskie, M. F., van Dyk, S., et al., 2003, The IRSA 2MASS All 
Sky Point Source Catalog, NASA/IPAC Infrared Science Archive,
http://irsa.ipac.caltech.edu/applications/Gator/\\   
Deharveng, L., Zavagno, A., Caplan, J. 2005, A\&A, 433, 565\\ 
Deharveng, L., Lefloch, B., Kurtz, S., Nadeau, D., Pomar\`{e}s, M., Caplan, J., Zavagno, A., 2008, A\&A, 482, 585\\
Devine, K. E., Churchwell, E. B., Indebetouw, R., Watson, C., Crawford, S. M., 2008, AJ, 135, 2095\\
Figuer\^{e}do, E., Blum, R. D., Damineli, A.,  Conti, P. S., 2002, AJ, 124, 2739\\
Georgelin, Y. M., Georgelin, Y. P., 1970, A\&A, 6, 349\\ 
Getman, K. V., Feigelson, E. D., Luhman, K. L., Sicilia-Aguilar, A., Wang, J.,  Garmire, G. P., 2009, ApJ, 699, 1454\\
Girardi, L., Bertelli, G., Bressan, A., Chiosi, C., Groenewegen, M. A. T., et al., 2002, A\&A, 391, 195\\ 
Haisch, K. E., Lada, E. A.,  Lada, C. J., 2000, AJ, 120, 1396\\ 
Haisch, K. E., Lada, E. A.,  Lada, C. J., 2001, AJ, 121, 2065\\ 
Hern\'{a}ndez, J., Hartmann, L., Megeath, T., Gutermuth, R., Muzerolle, J., 2007, ApJ, 662, 1067\\ 
Hillenbrand, L. A., Strom, S. E., Vrba, F. J.,  Keene, J., 1992, ApJ, 397, 613\\ 
Hillenbrand, L. A., Massey, P., Strom, S. E., Merrill, K. M., 1993, AJ, 106, 1906 \\ 
Hubble E., 1922, ApJ, 56, 400\\
Hunter, T. R., Testi, L., Taylor, G. B., Tofani, G., Felli, M.,  Phillips, T. G., 1995, A\&A, 302, 249\\
Jacoby, G. H., Hunter, D. A.,  Christian, C. A., 1984, ApJS, 56, 257\\
Johnson, H. L.  Morgan, W. W., 1953, ApJ, 117, 313\\ 
Johnson, H. L., 1966, ARA\&A, 4, 193\\ 
Jones, B. F.,  Herbig, G. H., 1979, 84, 1872\\
Jose, J., et al., 2008, MNRAS, 384, 1675\\ 
King, I., 1962, AJ, 67, 471\\ 
Kroupa, P., 2002, SCIENCE, 295, 82\\ 
Kroupa P., 2008, in Knapen J. H., Mahoney T. J., Vazdekis A., eds, ASP  Conf. Ser. Vol. 390, Pathways Through an Eclectic Universe. Astron.   Soc. Pac., San Francisco, p. 3\\
Lada, C. J.,  Lada, E. A., 1991, in ASP Conf. Ser. 13, The Formation and Evolution of Star Clusters, ed. K. Janes (San Francisco: ASP), 3\\ 
Lada, C. J.,  Adams, F. C., 1992, ApJ, 393, 278\\ 
Lada, C. J., Young, T.,  Greene, T., 1993, ApJ, 408, 471\\ 
Lada, E . A .  Lada, C . J., 1995, AJ , 109, 1682\\ 
Lada, C. J. et al., 2000, AJ, 120, 3162\\ 
Lada, C. J.,  Lada E. A., 2003, ARA\&A, 41, 57\\
Landolt A.U., 1992, AJ, 104, 340\\  
Larson, R. B., 1992, MNRAS, 256, 641\\
Leisawitz D., Bash F. N., Thaddeus P., 1989, ApJS, 70, 731\\ 
Leistra, A., Cotera, A. S.,  Liebert, J., 2006, AJ, 131, 2571\\
Leistra, A., Cotera, A. S., Liebert, J.,  Burton, M., 2005, AJ, 130, 1719\\
Mart\'{i}n-Hern\'{a}ndez, N. L., van der Hulst, J. M.,  Tielens, A. G. G. M., 2003,  A\&A, 407, 957\\
Martins, F., Schaerer, D., Hillier, D. J., 2005, A\&A, 436, 1049\\ 
Meyer, M., Calvet, N.,  Hillenbrand, L. A., 1997, AJ, 114, 288\\ 
Meyer, M. R., Adams, F. C., Hillenbrand, L. A., Carpenter, J. M.,  Larson, R. B., 2000, Protostars and Planets IV, 121\\
Meynet, G.,  Maeder, A., 2005, A\&A, 429, 581\\
Moffat, A. F. J.,  Fitzgerald, M. P., Jackson, P. D.,  1979, A\&AS, 38, 197\\
Muench, A. A., Lada, E.A.,  Lada, C.J., 2000, ApJ, 553, 338\\
Muench, A. A. et al., 2003, AJ, 125, 2029
Ogura K., Chauhan N., Pandey A.K., Bhatt B.C., Ojha D.K., Itoh Y., 2007, PASJ, 59, 199 \\
Ojha, D. K., Tamura, M., Nakijama, Y., et al., 2004a, ApJ, 608, 797\\ 
Ojha, D. K., Tamura, M., Nakajima, Y., et al, 2004b, ApJ, 616, 1042\\
Oliveira, J. M., Jeffries, R. D., van Loon, J. Th., Littlefair, S. P.,  Naylor, T., 2005, MNRAS, 358, L21\\
Pandey, A. K., Ogura, K.,  Sekiguchi, K., 2000, PASJ, 52, 847\\
Pandey A.K., Nilakshi, Ogura K., Sagar R.,  Tarusawa K., 2001, A\&A, 374, 504\\  
Pandey, A. K., Upadhyay, K., Nakada, Y.,  Ogura, K., 2003, A\&A, 397, 191\\ 
Pandey, A. K., Upadhyay, K., Ogura, K., Sagar, R., Mohan, V. et al., 2005, MNRAS, 358, 1290\\ 
Pandey, A. K., Sharma, S., Ogura, K., Ojha, D. K., Chen, W. P. et al., 2008, MNRAS, 383, 1241\\ 
Pomar\`{e}s, M., Zavagno, A., Deharveng, L., Cunningham, M., Jones, P., Kurtz, S. et al., 2009, A\&A, 494, 987\\
Price, N. M., Podsiadlowski, Ph., 1995, MNRAS, 273, 1041\\
Puga, E., Hony, S., Neiner, C., Lenorzer, A., Hubert, A.-M. et al., 2009, A\&A, 503, 107\\ 
Robin, A. C., Reyle, C., Derriere, S.,  Picaud, S., 2003, A\&A, 409, 523\\ 
Robitaille, T. P., Whitney, B. A., Indebetouw, R., Wood, K., Denzmore, P., 2006, ApJS, 167, 256\\
Salpeter, E.E., 1955, ApJ, 121, 161\\ 
Sanchawala, K. et al.,  2007, ApJ, 667, 963\\	
Schmidt-Kaler, Th. 1982, Landolt-Bornstein, Vol. 2b, ed. K. Schaifers, H. H. Voigt, H. Landolt (Berlin: Springer), 19\\
Sharma, S., Pandey, A. K., Ojha, D. K., Chen, W. P., Ghosh, S. K., Bhatt, B. C., Maheswar, G., Sagar, R., 2007, MNRAS, 380, 1141\\ 
Sharpless, S., 1959, ApJS, 4, 257\\
Siess, L., Dufour, E.,  Forestini, M., 2000, A\&A, 358, 593\\ 
Stetson, P. B., 1987, PASP, 99, 191\\  
Sugitani, K. et al., 2002, ApJ, 565, L25\\
Swarup, G., Ananthkrishnan, S., Kaphi, V. K., Rao, A. P., Subrhmanya, C. R.,  Kulkarni, V. K., 1991, Current Science, 60, 95\\
Tapia, M., Persi, P., Bohigas, J., Ferrari-Toniolo, M., 1997, AJ, 113, 1769\\ 
Teixeira, P. S., Fernandes, S. R., Alves, J. F., Correia, J. C., Santos, F. D., Lada, E. A.,  Lada, C. J., 2004, A\&A, 413, L1\\
Torres-Dodgen, Ana V., Weaver, W. B., 1993, PASP, 105, 693\\
Vacca, W. D., Garmany, C. D.,  Shull, J. M., 1996, A\&A, 460, 914\\
Walborn, N. R., Fitzpatrick, E. L., 1990, PASP, 102, 379\\
Wood, D. O. S.,  Churchwell, E., 1989, 340, 265\\ 
Yasui, C., Kobayashi, N., Tokunaga, A. T., Saito, M.,  Tokoku, C., 2009, ApJ, 705, 54\\
Zavagno, A., Pomar\`{e}s, M., Deharveng, L., Hosokawa, T., Russeil, D., Caplan, J., 2007, A\&A, 472, 835\\
Zinnecker, H., 1986., IMF in starburst regions. In light on Dark Matter, ed. F.P.Israel, ApSS Library Vol. 124, pp.277-278\\
Zinnecker, H., McCaughrean, M. J.,  Wilking, B. A., 1993, in Protostars and Planets III, ed. E. Levy \& J. Lunine (Tucson: Univ. Arizona Press), 429\\
                              
\begin{table}
\caption{Log of observations}
\label{obslog}

\begin{tabular}{p{.3in}p{.45in}p{.45in}p{.35in}p{1in}}
\hline $\alpha_{(2000)}$ & $\delta_{(2000)}$ & Date of &Filter &
Exposure time\\ (h:m:s) & ($^{\circ}:^{\prime}:^{\prime\prime}$)  &
observation  &   &  (s)$\times$no. of frames \\ \hline {\it HCT$^1$} &
&   &\\
  
04:40:38& +50:27:36& 2004.11.03 & $U$ &  600$\times$3 \\ 
04:40:38&+50:27:36& 2004.11.03 & $B$ &  300$\times$3, 60$\times$1, 20$\times$1\\ 
04:40:38& +50:27:36& 2004.11.03 & $V$ &  120$\times$3, 10$\times$1\\ 
04:40:38& +50:27:36& 2004.11.03 & $R$ &  60$\times$3, 10$\times$1\\ 
04:40:38& +50:27:36& 2004.11.03 & $I$ &  60$\times$3, 10$\times$1, 5$\times$1  \\ 
04:40:38& +50:27:36& 2007.01.26 & \sii&450$\times$1\\ 
04:40:38& +50:27:36& 2007.01.26 & \oiii&450$\times$1\\ 
04:40:37& +50:27:41& 2006.09.08 & Gr7/167l&900$\times$1 \\ 
04:40:39& +50:27:18& 2007.01.26 & Gr7/167l&600$\times$1\\ 
04:40:35& +50:28:44& 2007.01.26 & Gr7/167l&750$\times$1\\ 
04:40:32& +50:27:54& 2007.01.26 & Gr7/167l&750$\times$1\\ 
{\it ST$^2$} &                  &            &\\
04:40:38& +50:27:36& 2006.12.12 & $V$ &  300$\times$10 \\ 
04:40:38&+50:27:36& 2006.12.12 & $I_c$ &  300$\times$5 \\

\hline
\end{tabular} 

$^1$  2-m Himalayan Chandra Telescope, IAO, Hanle\\ $^2$  104-cm
Sampurnanand Telescope, ARIES, Naini Tal\\

\end{table}









\begin{table}

\caption{$UBVRI_cJHK$ photometric data of sample stars. The complete
table is available in electronic form only.}
\label{optdata}

\scriptsize
\begin{tabular}{cccccccccccc}

 \hline

star& $\alpha_{(2000)}$& $\delta_{(2000)}$& $V$ &$(U-B)$ &$(B-V)$ & $(V-R)$ & $(V-I)$ & $J$  & $H$  & $ K$  &  $A_V$\\
ID& (h:m:s)  &  ($^{\circ}:^{\prime}:^{\prime\prime}$)   &     &        &        &         &         &    &    &     &    \\
\hline
1& 04:39:46.271 & +50:30:00.70 &    18.415    &  -   &  - &  -  &  1.611  &  - &  - & - &  - \\
2& 04:39:46.320 & +50:22:03.89 &    21.320    &  -   &  - &  -  &  1.865  &  - &  - & - &  - \\
3& 04:39:46.320 & +50:22:23.00 &    20.893    &  -   &  - &  -  &  1.887  &  - &  - & - &  - \\
...&.....&.....&.....&.....&.....&.....&.....&.....&......&.....&\\
...&.....&.....&.....&.....&.....&.....&.....&.....&......&.....&\\

1155&04:40:32.181&+50:27:53.40&13.067&0.396 &    0.917&     0.542 &    1.055&    11.172 &10.838& 10.728 &3.1*\\
...&.....&.....&.....&.....&.....&.....&.....&.....&......&.....&\\
\hline       
\end{tabular}\\
$A_V$ for the $\star$ marked sources have been obtained using  optical photometry\\

\end{table}

\begin{table}
\caption{Completeness Factor of photometric data in the cluster
and field regions.}
\label{cf_opt}
\begin{tabular}{ccc} \hline
V range&    NGC 1624 &   Field region\\ (mag)& $r < 2^{\prime}$ & r
 $\ge$ $3^{\prime}$   \\ \hline 11 - 12&1.00&1.00\\ 12 -
 13&1.00&1.00\\ 13 - 14&1.00&1.00\\ 14 - 15&1.00&1.00\\ 15 -
 16&1.00&1.00\\ 16 - 17&0.98&0.98\\ 17 - 18&0.98&0.97\\ 18 -
 19&0.90&0.95\\ 19 - 20&0.90&0.93\\ 20 - 21&0.80&0.89\\ 21 -
 22&0.55&0.61\\

\hline
\end{tabular}
\end{table}


\clearpage

\newpage 

\begin{figure*} \centering \includegraphics[scale = 0.7,
trim = 10 30 10 120, clip]{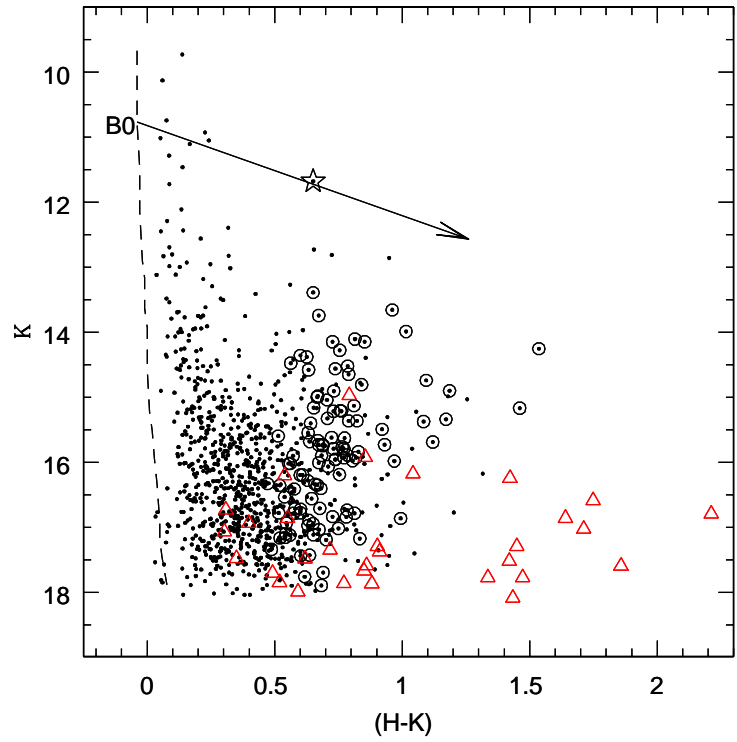}

\caption{$K/(H - K)$ CMD for the  sources detected in the  $J H K$ bands and having error $\le$ 0.15 mag.   
The encircled are the  NIR excess sources in the region and the  red triangles are the J drop out sources.
The vertical dashed line represents the unreddened ZAMS  locus shifted for the  cluster 
distance. The slanting line traces the reddening  vector for the B0 spectral class with reddening  $A_V$ = 
15 mag. The star symbol represents candidate ionizing source of the UC\hii region.   }

\label{hkk}
\end{figure*}

\begin{figure*} \centering \includegraphics[scale = 0.85,
trim = 0 0 0 0, clip]{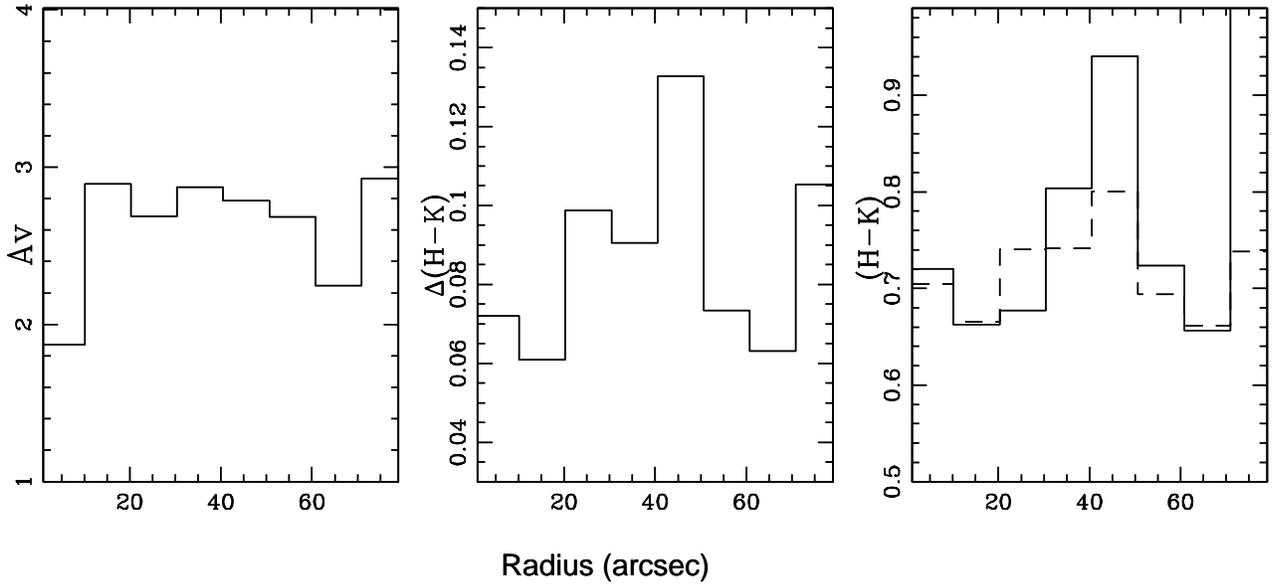}

\caption { {\it Left panel}: Radial variation of Av within a cluster radius of 80 arcsec.  
{\it Middle panel}: Radial variation of $\Delta(H-K)$,  defined as the horizontal  displacement 
from the reddening vector at the  boundary of `F' and `T' regions (see Fig. \ref{jhhk}) 
within a radius of 80 arcsec. {\it Right panel}: Radial variation of $(H-K)$ for the NIR excess 
sources (dashed histogram) and for all the sources detected in  $H$ and $K$- bands (solid histogram).  }

\label{radial}
\end{figure*}

\begin{figure*} 
\centering 
\includegraphics[scale = .5,  trim = 10 30 10 120, clip]{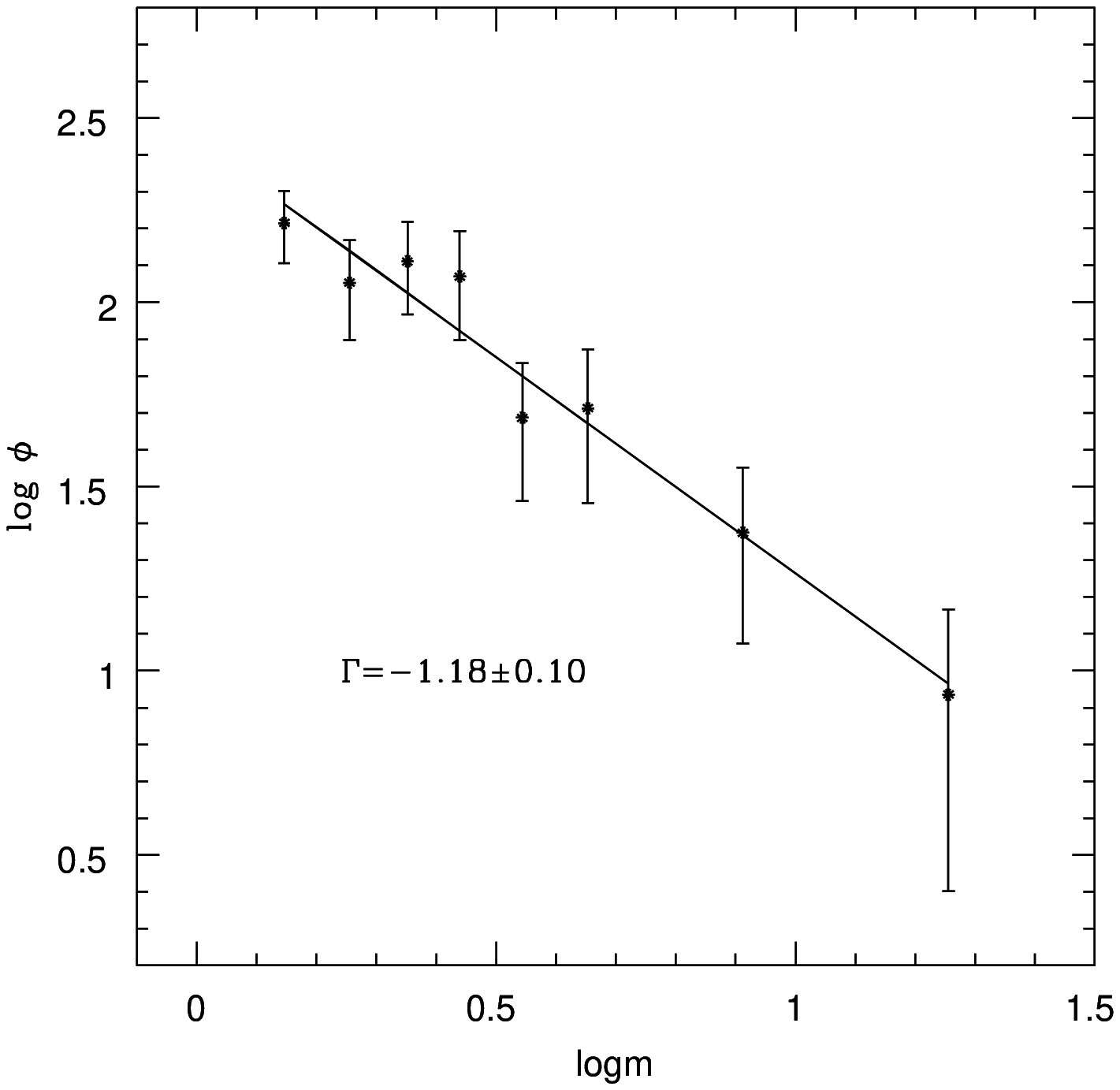}

\caption{A plot of the MF for NGC 1624 within  $r \le
2^\prime$ using optical data. The $\phi$ represents $N$/dlog $m$.  The error bars
represent  $\pm$$\sqrt{N}$ errors. The continuous line shows
least-squares fit  to the mass ranges described in the text.  The
value of the slope obtained is  mentioned in the figure. }
\label{mf}
\end{figure*}

\begin{figure*} \centering 
\includegraphics[scale = .5, trim = 10 20 10 120, clip]{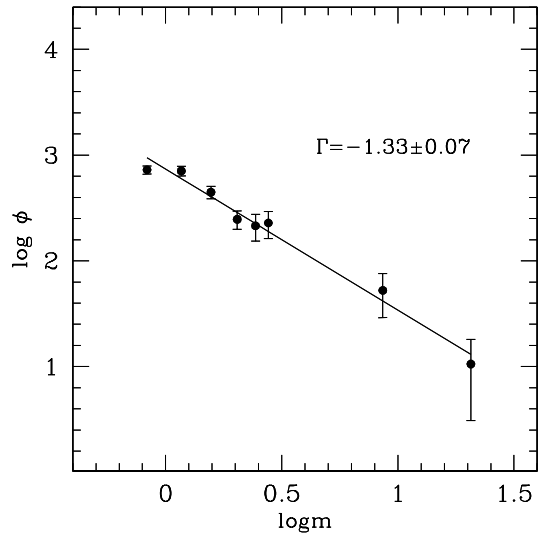}

\caption{A plot of the mass function for NGC 1624 within $\sim$ 9.6 arcmin$^2$ area
using the $J$-band data. The $\phi$ represents $N$/dlog $m$.  The error bars
represent  $\pm$$\sqrt{N}$ errors. The continuous lines show
least-squares fit  to the mass ranges described in the text.  The
value of the slope obtained is mentioned in the figure. }
\label{mf_ir}
\end{figure*}

\begin{figure*} \centering \includegraphics[scale = .5, trim = 10 30 10
120, clip]{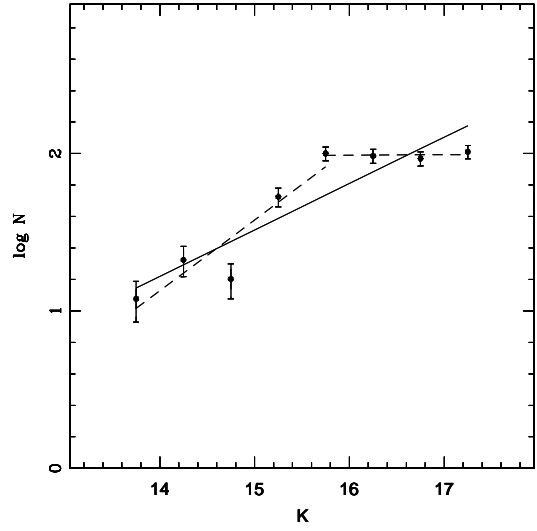}
\caption {KLF derived  after completeness correction and subtracting
the  field star contamination (see the text). The  linear  fit for various magnitude ranges are 
represented by the straight  lines.}
\label{klf}
\end{figure*}


\bsp

\end{document}